\documentclass[aps,prl,superscriptaddress,twocolumn, 10pt]{revtex4-1}
\usepackage{graphicx}
\usepackage{physics}
\usepackage{algpseudocode}
\usepackage{pythonhighlight}
\usepackage{amsmath,amssymb,amsfonts}
\usepackage{hyperref}
\usepackage{color}
\usepackage{natbib}
\usepackage{changes}
\usepackage[english]{babel}
\usepackage{subcaption}
\usepackage{subfiles}
\usepackage{float}

\begin{document}

\title{Efficient quantum algorithms for $GHZ$ and $W$ states, and implementation on the IBM quantum computer}
\author{Diogo Cruz}
\affiliation{These authors contributed equally to this work}
\affiliation{Institute of Physics, \'{E}cole Polytechnique F\'{e}d\'{e}rale de Lausanne (EPFL), Lausanne, CH-1015, Switzerland}
\author{Romain Fournier}
\affiliation{These authors contributed equally to this work}
\affiliation{Institute of Physics, \'{E}cole Polytechnique F\'{e}d\'{e}rale de Lausanne (EPFL), Lausanne, CH-1015, Switzerland}
\author{Fabien Gremion}
\affiliation{These authors contributed equally to this work}
\affiliation{Institute of Physics, \'{E}cole Polytechnique F\'{e}d\'{e}rale de Lausanne (EPFL), Lausanne, CH-1015, Switzerland}
\author{Alix Jeannerot}
\affiliation{These authors contributed equally to this work}
\affiliation{Institute of Physics, \'{E}cole Polytechnique F\'{e}d\'{e}rale de Lausanne (EPFL), Lausanne, CH-1015, Switzerland}
\author{Kenichi Komagata}
\affiliation{These authors contributed equally to this work}
\affiliation{Institute of Physics, \'{E}cole Polytechnique F\'{e}d\'{e}rale de Lausanne (EPFL), Lausanne, CH-1015, Switzerland}
\author{Tara Tosic}
\affiliation{These authors contributed equally to this work}
\affiliation{Institute of Physics, \'{E}cole Polytechnique F\'{e}d\'{e}rale de Lausanne (EPFL), Lausanne, CH-1015, Switzerland}
\author{Jarla Thiesbrummel}
\affiliation{These authors contributed equally to this work}
\affiliation{Institute of Physics, \'{E}cole Polytechnique F\'{e}d\'{e}rale de Lausanne (EPFL), Lausanne, CH-1015, Switzerland}
\author{Chun Lam Chan}
\author{Nicolas Macris}
\email{nicolas.macris@epfl.ch}
\affiliation{Laboratoire de Th\'{e}orie des Communications, Facult\'{e} Informatique et Communications, \'{E}cole Polytechnique F\'{e}d\'{e}rale de Lausanne (EPFL), Lausanne, CH-1015, Switzerland}
\author{Marc-Andr\'{e} Dupertuis}
\email{marc-andre.dupertuis@epfl.ch}
\affiliation{Institute of Physics, \'{E}cole Polytechnique F\'{e}d\'{e}rale de Lausanne (EPFL), Lausanne, CH-1015, Switzerland}
\author{Cl\'{e}ment Javerzac-Galy}
\email{clement.javerzac-galy@epfl.ch}
\affiliation{Institute of Physics, \'{E}cole Polytechnique F\'{e}d\'{e}rale de Lausanne (EPFL), Lausanne, CH-1015, Switzerland}

\begin{abstract}
We propose efficient algorithms with logarithmic step complexities for the generation of entangled $GHZ_N$ and $W_N$ states useful for quantum networks, and we demonstrate an implementation on the IBM quantum computer up to $N=16$. Improved quality is then investigated using full quantum tomography for low-$N$ GHZ and W states. This is completed by parity oscillations and histogram distance for large $N$ GHZ and W states respectively. We are capable to robustly build states with about twice the number of quantum bits which were previously achieved. Finally we attempt quantum error correction on GHZ using recent schemes proposed in the literature, but with the present amount of decoherence they prove detrimental.
\end{abstract}

\maketitle

\section{Introduction}\label{introSec}

Quantum entanglement involves non-local correlations between subsystems which were central in the historic debate which revealed stunning and fundamental non-locality of physics. Nowadays quantum entanglement between quantum bits is no less essential as a fundamental resource in quantum information, from quantum network protocols to quantum computing. The IBMQ experimental quantum computer, freely accessible by Internet~\cite{ibm}, is a fundamental tool enabling democratic student training in quantum computing. On the IBMQ platform two recent experiments~\cite{Ozaeta2017,Wang2018} have proved on one hand that 8 qubit GHZ states could be realized and on the other hand that 16 qubits could be fully entangled. In the present paper we move forward by proposing new algorithms and by extending the experimental realization and characterization up to $N=16$.

Among multipartite quantum states of $N$ quantum bits, at least two classes play a particular role: 
\begin{eqnarray}
\ket{GHZ_N} &= &\frac{1}{\sqrt{2}} \left( \ket{0}^{\otimes N} + \ket{1}^{\otimes N} \right) \nonumber \\
\ket{W_N} \quad &= &\frac{1}{\sqrt{N}} \Big( \ket{100\ldots 0} + \ket{010\ldots 0} +  \nonumber \\
& & \qquad\qquad \cdots + \ket{000\ldots 1} \Big) \nonumber 
\end{eqnarray}
$GHZ_3$ states were originally proposed to falsify local realistic theories in one shot~\cite{Greenberger1990}, and their generalization for arbitrary $N$ are superposition states that can be viewed as the ``biggest Schr\"{o}dinger cat'' states for $N$ qubits, since the kets $\ket{0}^{\otimes N}$ and $\ket{1}^{\otimes N}$ are most far apart kets in Hilbert space. $W_N$ states are fundamentally different multipartite quantum states, as revealed by~\cite{Dur2000} who showed that for $N\geq 3$ these two states define two classes whose states cannot be converted into one another by LOCC (Local Operations and Classical Communications), and exhaust the possible purely tripartite classes for $N=3$. For larger $N$ the number of different classes increase rapidly so entanglement is now considered as a very rich and challenging concept, spurring a huge research effort~\cite{Horodecki2009,Vedral2014}. In the present paper, we concentrate on $GHZ_N$ and $W_N$ which are two typical representatives of the general multipartite entangled jungle. They also constitute very interesting and powerful targets as fundamental resources in distributed quantum information processing~\cite{Dondt2005,Yesilyurt2016} (see also the book reviews~\cite{Bengtsson2016,Walter2017}). The properties and potential of $GHZ_N$ and $W_N$ are completely different: the former are fragile against loss and important for example in quantum secret sharing, whilst the latter are robust against loss and central to quantum memories, multiparty quantum network protocols and universal quantum cloning machines. 
Many schemes have been designed to produce these states for large $N$ (see for example~\cite{Zang2017}), but it is also important to consider other aspects like probabilistic or deterministic creation, possibility to extend them locally in quantum networks~\cite{Yesilyurt2016}, or to transform them nonlocally~\cite{Tashima2016}.

In the first section we present the algorithms we developed to generate $GHZ_N$ and $W_N$, with emphasis on so-called logarithmic algorithms, which typically take $\sim\log_2N$ steps. For quantum networks these deterministic algorithms open the possibility to create a shared $GHZ_N$ or $W_N$ by local extension without prior entanglement, even when the number of parties is arbitrary. In the second section we shortly detail the implementation strategy on IBMQ, before characterizing in the next section the successfully created states with quantum tomography, parity oscillations decay, fidelity and histogram distance. The efficiency of the logarithmic algorithms is demonstrated up to $N=15$ or $N=16$, a technical limit posed by the IBMQ processors at our disposal. We evidence clearly the effect of decoherence. Finally in the last section~\ref{qerSec} an attempt at error correction is made using a recently proposed scheme. The results show that for the moment, and for our purposes, quantum error correction is still impractical.

\section{Linear and Logarithmic algorithms for GHZ and W states}\label{algoSec}

We discuss the circuits for the generation of $GHZ_N$ and $W_N$ states. We first present circuits of linear time-complexity and proceed to the construction of logarithmic time-complexity ones. While this is pretty straightforward for $GHZ_N$ it is less so in the case of $W_N$. In this section we only describe the basic theoretical circuits. The concrete implementation on IBM quantum computers involves some extra cost due to connection constraints between qubits, an issue that is discussed in the next section.



\subsection{Construction of $GHZ_N$}\label{ghzAlgoSubSec}

\begin{figure}[!hbtp]
\centering
\includegraphics[width=6cm]{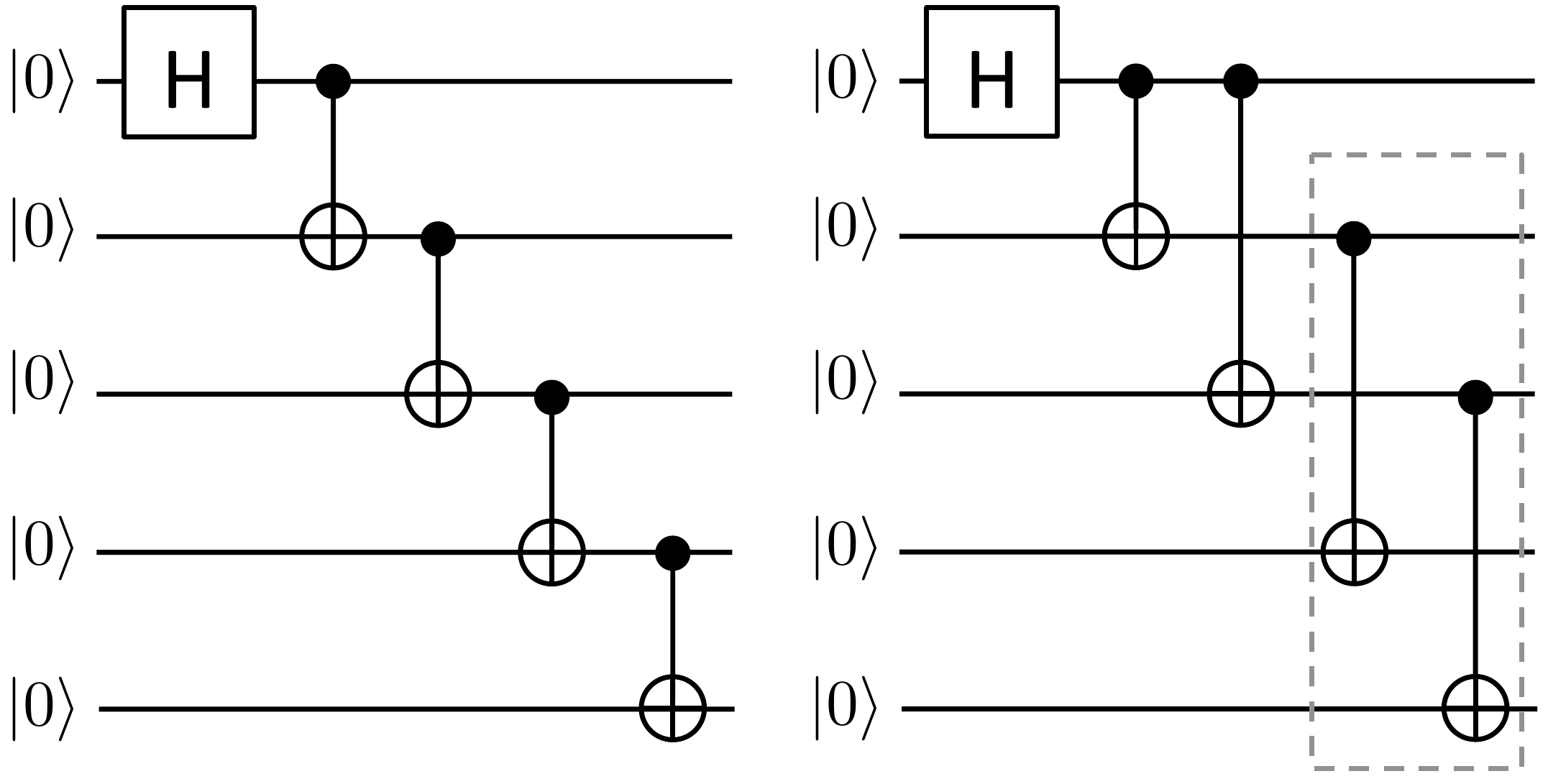}
\caption{Left: standard linear complexity circuit for $\ket{GHZ_5}$. Right: reshuffling control bits appropriately allows the last two CNOT gates to parallelize.}
\label{fig:GHZ-linear-N=5}
\end{figure}

A straightforward way to generate $GHZ_N$ states in linear time complexity $N$ is shown on the left Fig. \ref{fig:GHZ-linear-N=5}. This is the underlying circuit used on IBMQ in \cite{Ozaeta2017} for example, and would correspond in a quantum network to local linear extension with an ancilla qubit. It is not difficult to see that there is huge freedom in modifying the control bits of the CNOT gates while at the same time preserving the $\ket{GHZ_5}$ output state. The right hand side of Fig. \ref{fig:GHZ-linear-N=5} shows one such example\footnote{The part of this circuit for the first three qubits is essentially discussed in the documentation of IBMQ \cite{ibm} as a way to create
$\ket{GHZ_3}$}. The dotted lines indicate 'time slice' within which the CNOT gates can be performed in parallel (indeed one can move the CNOT gates along the lines as long as control bits are not exchanged with target bits and as long as target bits are not exchanged). The advantage of the right hand side is therefore that the number of 'time step' is $1+3=4$ instead of $1+4=5$ on the left hand side. 

This elementary remark is at the basis of an algorithm constructing circuits with logarithmic time complexity, and can also be practiced in a quantum network if the nodes act in parallel. For powers of two, $N= 2^m$, the following algorithm generates circuits of time complexity $1 + \log_2 N$. Note that all entry bits are initialized to $\ket{0}$. 

\vspace{0.05cm}

\begin{figure}[!hbtp]
\centering
\includegraphics[width=8.5cm]{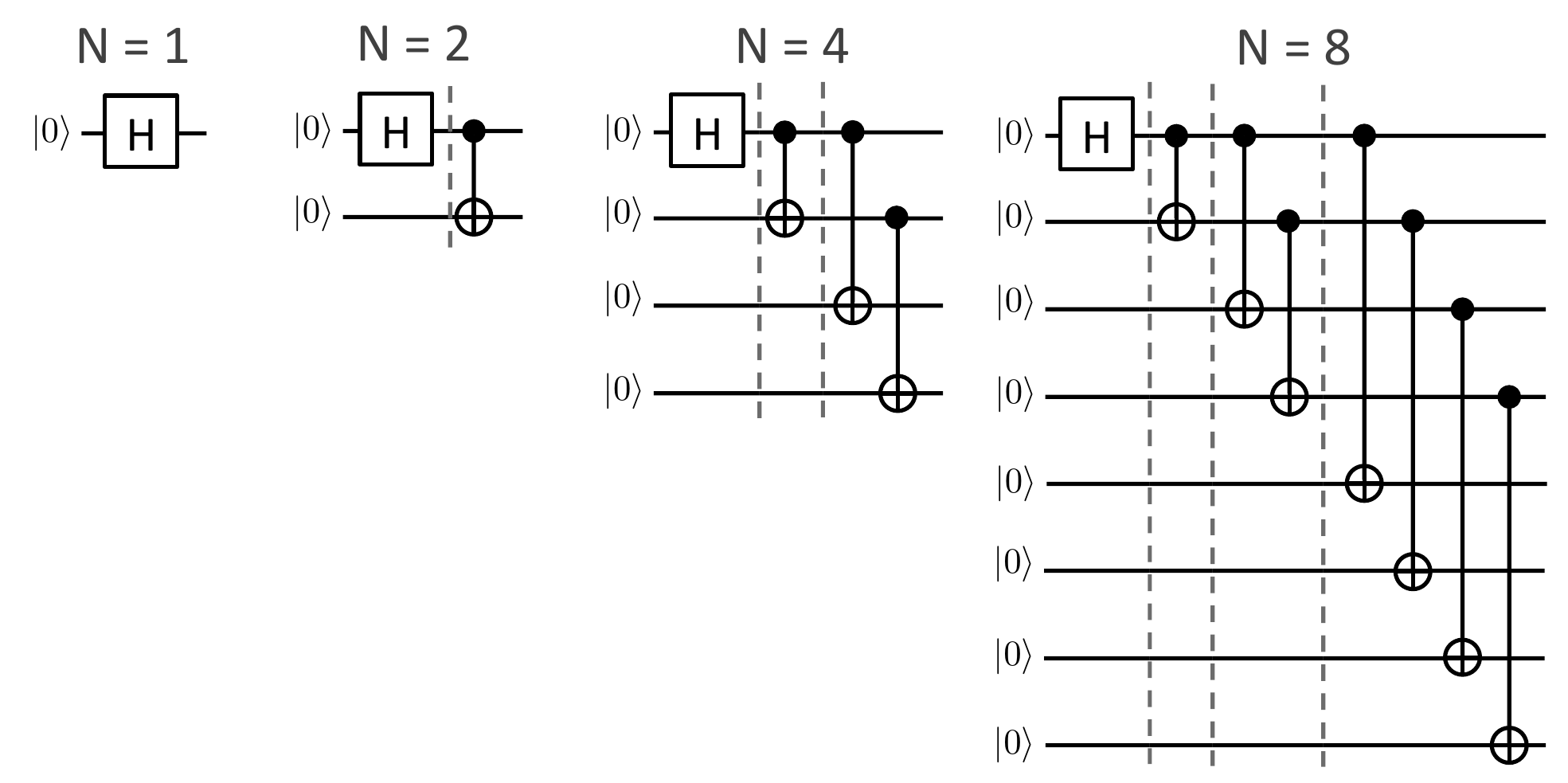}
\caption{Logarithmic time complexity circuits for $GHZ_N$ states with $N=1, 2, 4, 8$. Dotted lines separate time slices inside which CNOT's operate in parallel.}
\label{fig:GHZlogalgo}
\end{figure}

\begin{algorithmic}[1]
\Procedure{Create-GHZ-state}{$2^m$}
\If{m = 0}
	\State circuit $\gets$ connect a $H$ gate to the only qubit
   \Else
   	\State circuit $\gets$ CREATE-GHZ-STATE($2^{m-1}$)
    \For{every qubit $Q_i$ in the circuit}
    	\State Expand the circuit by connecting a new qubit to a CNOT gate with the controlled qubit $Q_i$
    \EndFor
\EndIf
\EndProcedure
\end{algorithmic}
This algorithm in action is shown for $N=1, 2, 4, 8$ on Fig. \ref{fig:GHZlogalgo}.

Finally, to obtain circuits for general $N$ not a power of two one may consider the circuit for $2^m$ immediately above $N$ and simply erase the last $2^m - N$ lines. 
For GHZ states there is a lot of freedom in order to parallelize the CNOT gates in a circuit, and the procedure shown here is not unique.
In general the circuit contains $N$ gates and has $\sim \log_2N$ time complexity.

\subsection{Construction of $W_N$}\label{wnAlgoSubSec}

In this paragraph we describe algorithms to construct circuits generating $W_N$ states in $\sim N$ and $O\sim \log_2 N$ time steps. To our knowledge the ideas presented here have not appeared in the literature for a general number of qubits. Low-complexity hierarchical schemes reminiscent of ours for $N=2^m$ have been discussed in~\cite{Yesilyurt2016} (and references therein) however not the case of general $N$. See also the special case $N=3$ on~\cite{phys-stackexchange-W} which seems however impractical as it uses a Toffoli gate.

\begin{figure}[!hbtp]
\centering
\includegraphics[width=3cm]{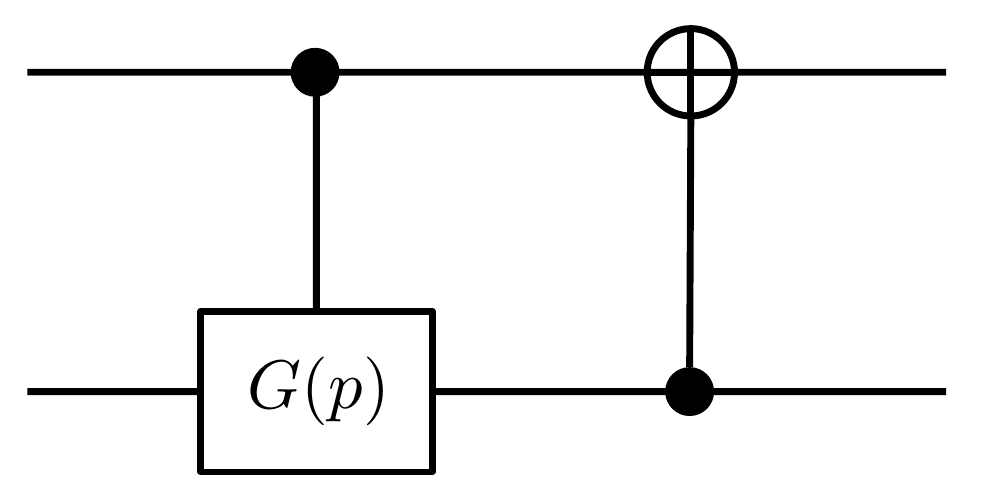}
\caption{Fundamental building block: $B(p)$ is  a controlled-$G(p)$ rotation followed by an inverted CNOT.}
\label{figs:buildingblock}
\end{figure}

Both constructions use the building block $B(p)$ with parameter $0 < p < 1$ in Fig. \ref{figs:buildingblock} where the rotation matrix is 
\begin{align*}
G(p) = \begin{pmatrix} \sqrt{p} & -\sqrt{1-p} \\ \sqrt{1-p} & \sqrt{p} \end{pmatrix}
\end{align*}
Note that for $p=1/2$ this is the transposed Hadamard gate.
In practice we will only use the following relations 
\begin{align*}
B(p) \ket{00} = \ket{00}, \quad B(p)\ket{10} = \sqrt{p}\ket{10} + \sqrt{1-p}\ket{01}
\end{align*}
Two concrete implementations of the controlled-$G(p)$ gate in terms of a usual rotation around the $y$-axis (available with the $U_3$ IBM gate) and the CNOT are discussed at the end of this section. 

\subsubsection{Linear time complexity circuit} 

\begin{figure}[!hbtp]
\centering
\includegraphics[width=7cm]{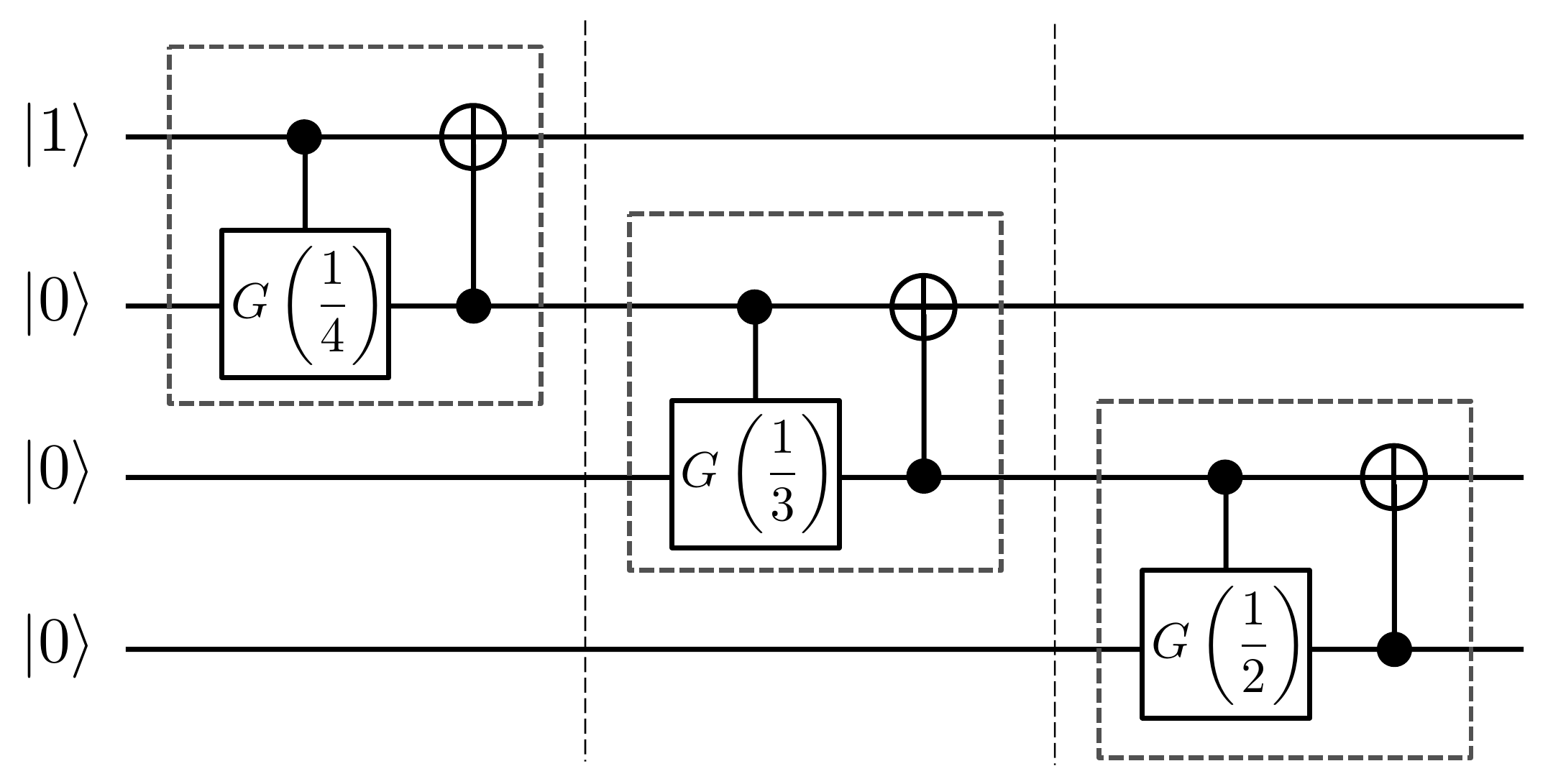}
\caption{A linear complexity circuit for $\ket{W_4}$.}
\label{figs:linearcircuit}
\end{figure}

This circuit is best described through a simple example for $N=4$ given on Fig. \ref{figs:linearcircuit}. 
It is perhaps instructive to briefly go through the calculation of the output. The initial state
is $\ket{1000}$. After the first $B(1/4)$ block it becomes $\sqrt{1/4}\ket{1000} +\sqrt{3/4}\ket{0100}$.
After the $B(1/3)$ block we find the intermediate state
\begin{align*}
\sqrt{\frac{1}{4}}\ket{1000} +\sqrt{\frac{3}{4}\frac{1}{3}}\ket{0100} + \sqrt{\frac{3}{4}\frac{2}{3}}\ket{0010} 
\end{align*}
and after the last block $B(1/2)$ we get the final state
\begin{align*}
\sqrt{\frac{1}{4}}\ket{1000} & +\sqrt{\frac{3}{4}\frac{1}{3}}\ket{0100} 
+ 
\sqrt{\frac{3}{4}\frac{2}{3}\frac{1}{2}}\ket{0010}
\nonumber \\ &
+  
\sqrt{\frac{3}{4}\frac{2}{3}\frac{1}{2}}\ket{0001} = \ket{W_4}
\end{align*}

We remark that the CNOT operations could have been performed at the very end 
of the circuit. However for implementations on IBMQ this has the disadvantage that the intermediate states contain many qubits in the $\ket{1}$ state which has a tendency to relax to the ground state
$\ket{0}$ thereby introducing unwanted errors. The present form of the circuit minimizes the number of $\ket{1}$ states at any given time (between $G$ and $CNOT$ gates we get two $1$'s and this is the maximal number of $1$'s at any given time). 

It is quite clear that the general circuit for constructing $\ket{W_N}$ will involve the sequence of blocks $B(1/N), B(1/(N-1)), \cdots, B(1/3), B(1/2)$. The total number of such gates 
and time steps are $N-1$. Counting the number of gates involved in the realization of 
$B(p)$, depending on the two ways the controlled-$G(p)$ gate is realized, multiplies this complexity by a factor $5$ or $4$. In any case the overall number of gates is $\sim N$ and the time complexity is also $\sim N$.

\subsubsection{\it Logarithmic time complexity circuit} 

\begin{figure}[!hbtp]
\centering
\includegraphics[width=8cm]{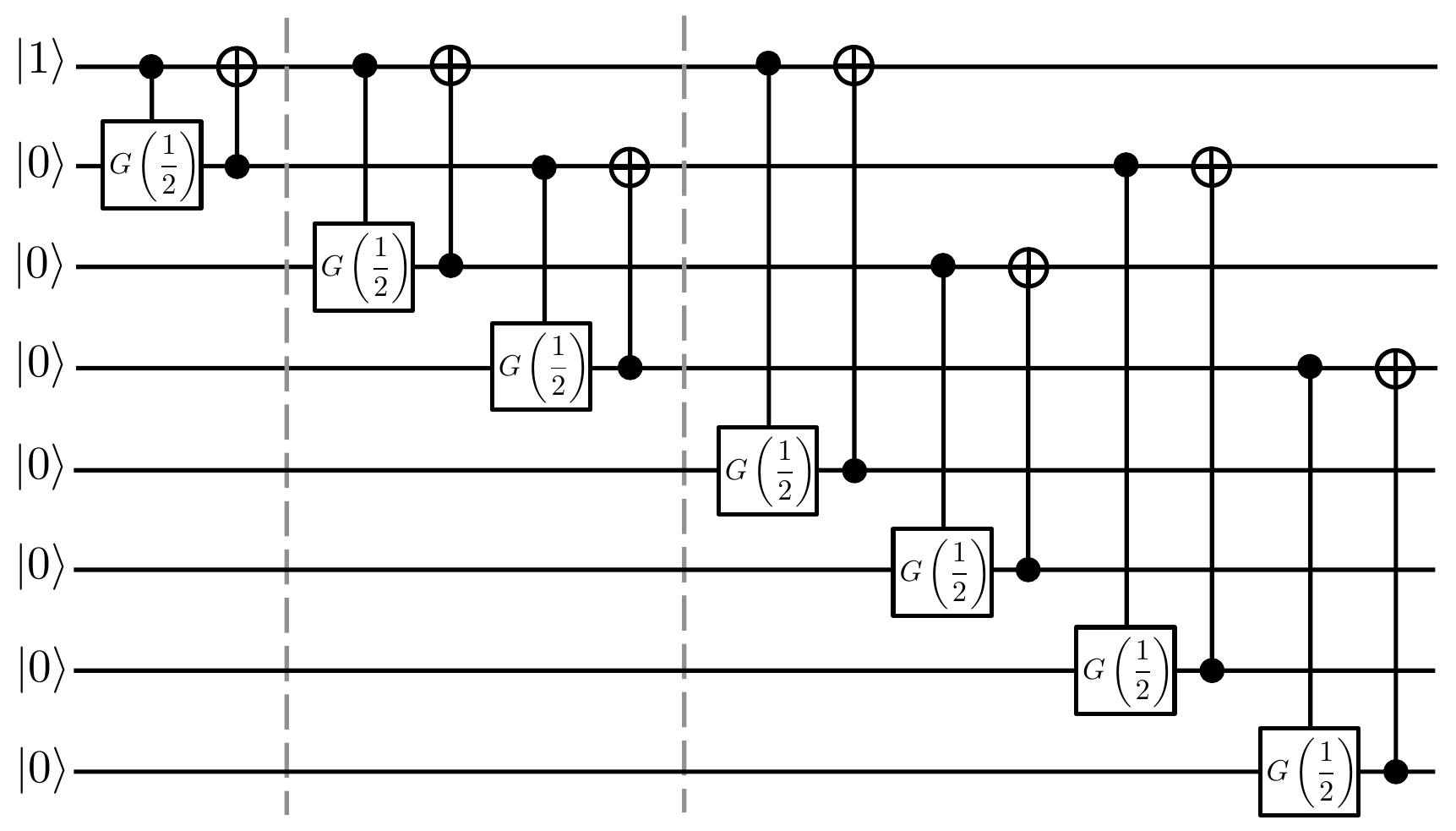}
\caption{Logarithmic complexity circuit for $\ket{W_8}$.}
\label{figs:logWN=8}
\end{figure}

We start with the particularly transparent special case when the number of qubits is a power of two, namely $N = 2^m$.
The circuit construction is illustrated for $N=8$ on Fig. \ref{figs:logWN=8} and the generalization to arbitrary powers of two is manifest. 
\begin{figure}[!hbtp]
\centering
\includegraphics[width=3cm]{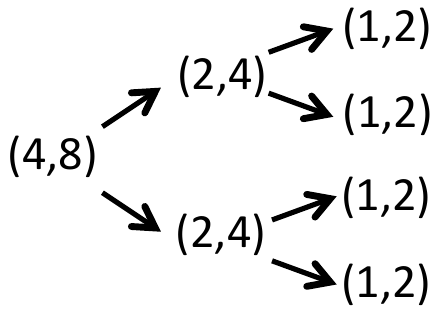}
\caption{Dichotomy tree for the construction of $\ket{W_8}$.}
\label{figs:logW-tree-N=8}
\end{figure}

\begin{figure}[!hbtp]
\centering
\includegraphics[width=6cm]{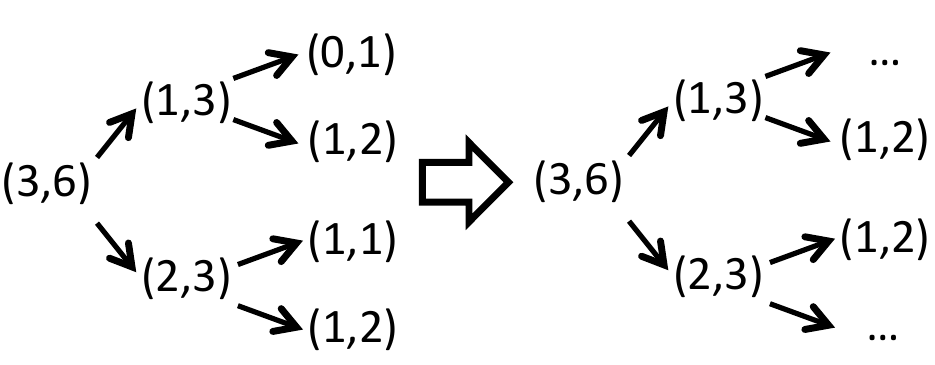}
\caption{Dichotomy tree for the construction of $\ket{W_6}$. Leaf nodes $(0,1)$ and $(1,1)$ are pruned and $(1,2)$ takes the place of $(1,1)$.}
\label{figs:logW-tree-N=6}
\end{figure}
\begin{figure}[!hbtp]
\centering
\includegraphics[width=6cm]{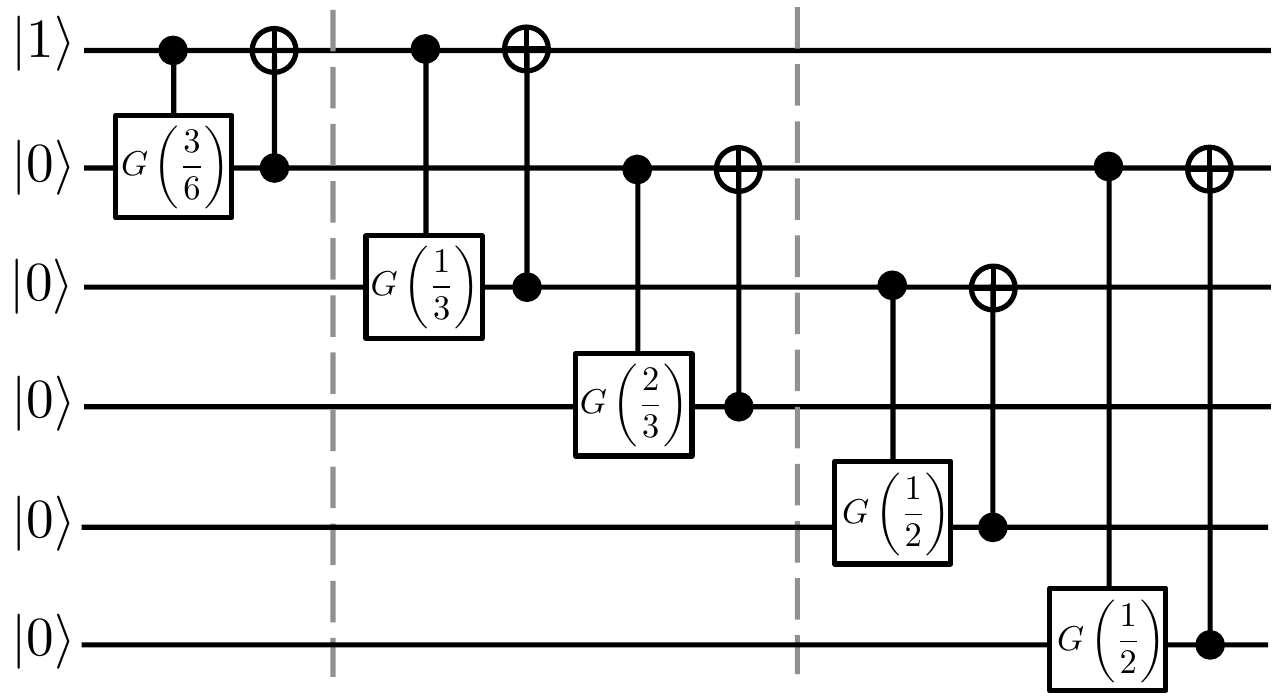}
\caption{Logarithmic complexity circuit for $\ket{W_6}$.}
\label{figs:logWN=6}
\end{figure}
The circuit is particularly simple as it involves only the controlled-$G(1/2)$ gate (basically equivalent to a controlled Hadamard) and CNOT. For the entry of the circuit, the top qubit is set to $\ket{1}$ while all other qubits are initialized to $\ket{0}$. The dashed lines indicate separations between ``time-slices", where within a time-slice $B(1/2)$ operations can be performed {\it in parallel}. As a result the time complexity is here $3k$ where $k= 5$ or $k=4$ depending on the implementation of the controlled-$G(1/2)$. In general the time complexity is $k\log_2 N$ and the total number of gates in the circuit is $k N$.

Let us highlight the tree-like structure behind the circuit 
of Fig. \ref{figs:logWN=8}. The first block $B(1/2) = B(4/8)$ in the first ``time-slice" gives birth to two children blocks $B(1/2) = B(2/4)$ that act in parallel in the second ``time-slice". Each child in the second ``time-slice" gives birth to two other children $B(1/2) = B(1/2)$ so that there are $8$ children in the third ``time-slice". We see that the construction of this tree explicitly shown in Fig.
\ref{figs:logW-tree-N=8}
proceed with perfect dichotomies 
of the powers of two. 

Let us now turn to the more challenging case of arbitrary $N$. When $N$ is not a power of two it is not possible to perform only perfect dichotomies but one can proceed with the ``best possible" dichotomies. By definition a best possible dichotomy of an integer $n$ is $n = \lfloor n/2 \rfloor + \lceil n/2 \rceil$ where $\lfloor n/2 \rfloor$, $\lceil n/2 \rceil$ are the lower and upper integer parts of $n/2$. The main idea is to first build a {\it tree of dichotomies} from which we read out the circuit in a completely systematic fashion. The algorithm is formalized in the Appendix.

Here we illustrate the main ideas behind the algorithm with the example of $N=6$ (Figs. \ref{figs:logW-tree-N=6}, \ref{figs:logWN=6}). 
First we consider the (here perfect) dichotomy $6 = 3 +3$ and retain the smallest integer (here $3$) and form the couple $(3,6)$. This is the root of the tree. The root generates two children as follows. It performs the two dichotomies $3=1 +2$ and $6=3+3$ and generate the upper child $(1,3)$ and lower child $(2,3)$. Each child generates two children by the same process. Child $(1,3)$ performs the dichotomies 
$1=0+1$ and $3=1+2$ and generates the upper child $(0,1)$ and lower child $(1,2)$. The child $(2,3)$ performs the dichotomies $2=1+1$ and $3=1+2$ and generates the children $(1,1)$ and $(1,2)$. The tree of dichotomies is shown on the left of Fig. \ref{figs:logW-tree-N=6}. In the last generation there are always only children of three types: $(0,1)$, (1,1) and $(1,2)$. It turns out that we must: (i) exchange $(1,1)$ with $(1,2)$ and then (ii) eliminate $(0,1)$ and $(1,1)$. This yields the final tree on the right in Fig. \ref{figs:logW-tree-N=6}.

The circuit of Fig. \ref{figs:logWN=6} is then built by connecting the successive blocks according to the tree structure. Each node $(n_1, n_2)$ in the tree corresponds to a block $B(n_1/n_2)$. The connections at each generation are as follows. For an ``upper child" the controlled rotation is controlled by the upper line of the parent. For a ``lower child" the controlled rotation is controlled by the lower line of the parent.

It is a good exercise to check the computational steps that lead to the output of Fig. \ref{figs:logWN=6} is $\ket{W_6}$. We give here the intermediate states at the end of each time slice and leave the details of calculations within as an exercise for the reader.
At the end of the first time slice (once the block $B(3/6)$ has acted) the state is 
\begin{align*}
\sqrt{\frac{3}{6}}\ket{100000} + \sqrt{\frac{3}{6}}\ket{010000}
\end{align*}
At the end of the second time slice once the two blocks $B(1/3)$, $B(2/3)$ have acted (in parallel) the state is
\begin{align*}
&\sqrt{\frac{3}{6}\frac{1}{3}}\ket{100000} 
+ \sqrt{\frac{3}{6}\frac{2}{3}}\ket{001000}
\nonumber \\ &
+ \sqrt{\frac{3}{6}\frac{2}{3}}\ket{010000}
+\sqrt{\frac{3}{6}\frac{1}{3}}\ket{000100}
\end{align*}
Finally at the output once the last two blocks have acted (in parallel) we find
\begin{align*}
&\sqrt{\frac{3}{6}\frac{1}{3}}\ket{100000} 
+ \sqrt{\frac{3}{6}\frac{2}{3}\frac{1}{2}}\ket{001000}
+ \sqrt{\frac{3}{6}\frac{2}{3}\frac{1}{2}}\ket{000010}
\nonumber \\ &
+ \sqrt{\frac{3}{6}\frac{2}{3}\frac{1}{2}}\ket{010000}
+ \sqrt{\frac{3}{6}\frac{2}{3}\frac{1}{2}}\ket{000001}
+\sqrt{\frac{3}{6}\frac{1}{3}}\ket{000100}
\end{align*}
which is nothing else than $\ket{W_6}$. 

In the Appendix the algorithm is formalized and also applied to generate a circuit for $N=11$. 

\subsubsection{The controlled-$G(p)$ gate.} 

We have at our disposal the $U_3(\theta, \phi, \lambda)$ IBM gate:
\begin{align*}
U_3(\theta, \lambda, \phi) 
= 
\begin{pmatrix} 
\cos\frac{\theta}{2}  &  - e^{-i\lambda} \sin\frac{\theta}{2} \\
e^{i\phi} \sin\frac{\theta}{2}  &  e^{i(\lambda + \phi)}\cos\frac{\theta}{2}
\end{pmatrix}
\end{align*}
For $\phi=\lambda=0$ which amounts to a standard $SU(2)$ representation of a rotation of angle $\theta$ around the $y$-axis. 
The angle $\theta$ is related to $p$
through $\cos(\theta/2) =\sqrt{p}$ and $\sin(\theta/2) = \sqrt{1-p}$. Our task here is to implement this rotation in a {\it controlled} way using {\it only} $U_3(\theta, 0, 0)$ and $CNOT$. It is well known that there are various ways of implementing controlled single qubit gates thanks to the pure single qubit gate and CNOT (see for example~\cite{nielsen_chuang},
Chapter 4). The following two identities using respectively $4$ and $3$ gates do the job:
\begin{align*}
& CG(p) \!=\! I\otimes U_3(\frac{\theta}{2}, 0,0) \,\, CNOT\,\, I\otimes U_3(-\frac{\theta}{2},0,0) \,\, CNOT\\
\\ &
CG(p) \!= \! I\otimes U_3(-\theta^\prime, 0, 0) \,\, CNOT \,\, I\otimes U_3(\theta^\prime, 0, 0)
\end{align*}
where $\sin \theta^\prime = \cos(\theta/2)$. In both identities, the first qubit is the control bit and the second is the target bit.

\section{Implementation}\label{implStratSec}

We can rename the qubits as we wish, to fit the architecture of the physical device, and the order of magnitude of the complexity won't be affected.
However, we can't always reach a minimal complexity for various reasons, some of which are briefly discussed below. 

{\it GHZ circuits.} The first used qubit requires $\sim\log_2(N)$ CNOT connections to other qubits, for the algorithm to be correctly implemented. The other qubits must also have a minimum number of connections. If this is not possible (for the 16-qubit device this is the case when $N\geq 13$) then it is necessary to distribute new CNOT gates through extra steps, using qubits that still have unused connections. Unfortunately this increases the total number of steps slightly. For example, for the 16-qubit case the number of steps increases from 5 to 6. There is also the issue of CNOT gate direction: sometimes the CNOT gate must be inverted, due to restrictions of the architecture, which will further increase the number of steps. In the 16-qubit case, the increase is from 6 to 10 steps.

{\it W circuits.} In practice we can remove the control in the rotation of the {\it first} block and leave the rotation always activated. This is so because the top-most qubit is always set to $\ket{1}$. Therefore (for $N=2^m$) the implemented version requires  $4N-5$ gates and $4\log_2(N)-2$ steps, to produce a $W_N$ state.
Still, there are issues similar to the ones just discussed above: the first used qubit requires $\sim\log_2(N)$ CNOT connections to other qubits and the other qubits must also have a minimum number of connections for the algorithm to function properly. If this is not possible (for the 16-qubit device, this is the case when $N\geq 13$) then we must distribute new CNOT gates and unfortunately this slightly increases the total number of steps. Again there is also the issue of CNOT gate direction. In the 16-qubit case, when all is said and done, the number of steps increases from 14 to 17. Finally, throughout all this analysis the $U_3$ gate was considered to take one time step. However, IBM states that this gate takes two time steps, so in practice the overall number of time steps for the W state algorithm will slightly increase (but remains $\sim\log_2 N$).

\begin{figure*}[!hbtp]
	\centering
	\includegraphics[width=\linewidth]{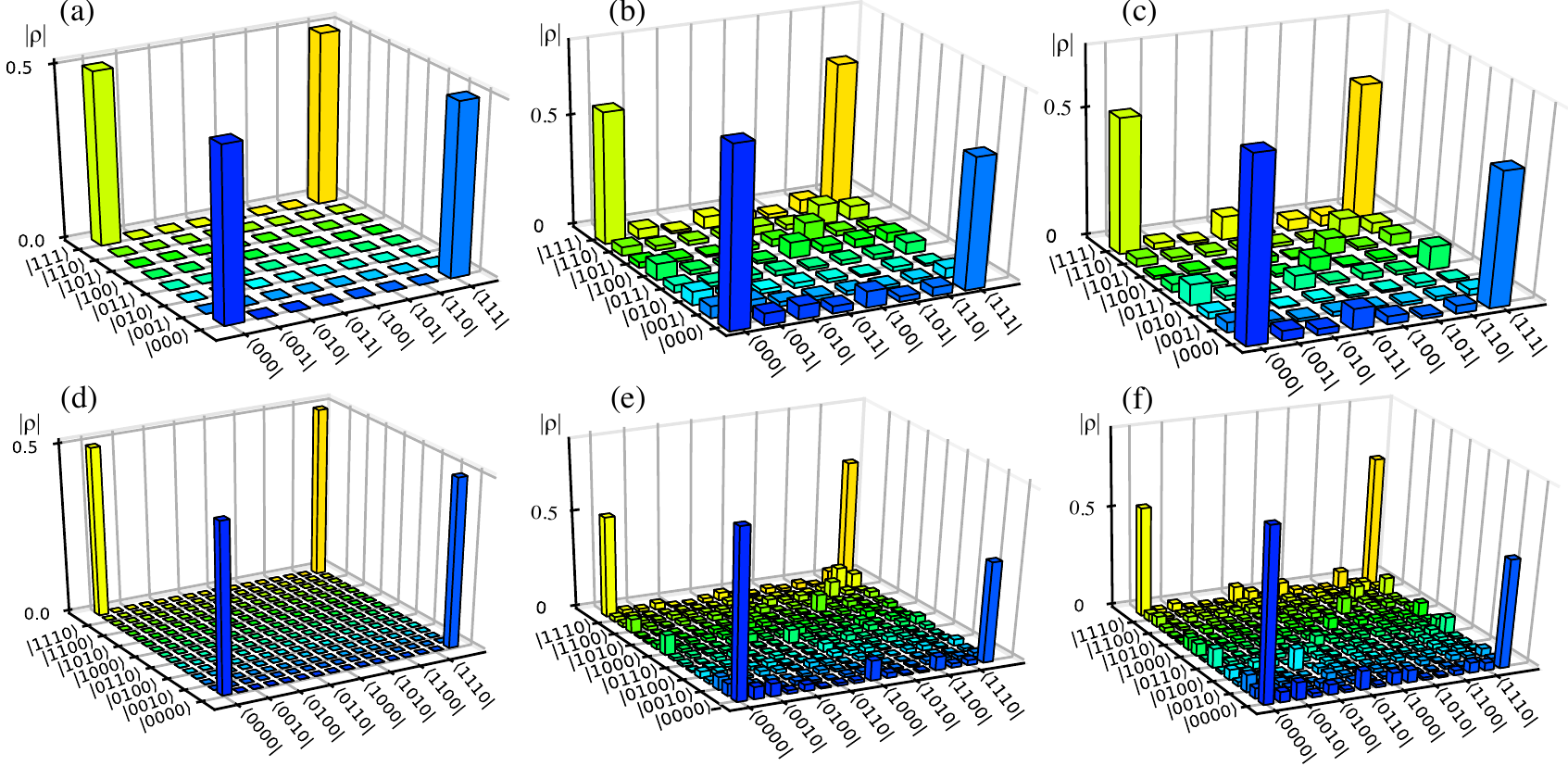}
    \caption{Absolute value of reconstructed density matrix elements of $GHZ_3$ states (upper row) and $GHZ_4$ (lower row). In the first column the ideal density matrix obtained with IBM quantum simulator is displayed, in the second column the measured tomography generated by the linear algorithm, and in the third column the same for the logarithmic algorithm.}
    \label{fig:GHZ_tomo}
\end{figure*}

\begin{figure*}[!hbtp]
	\centering
	\includegraphics[width=\linewidth]{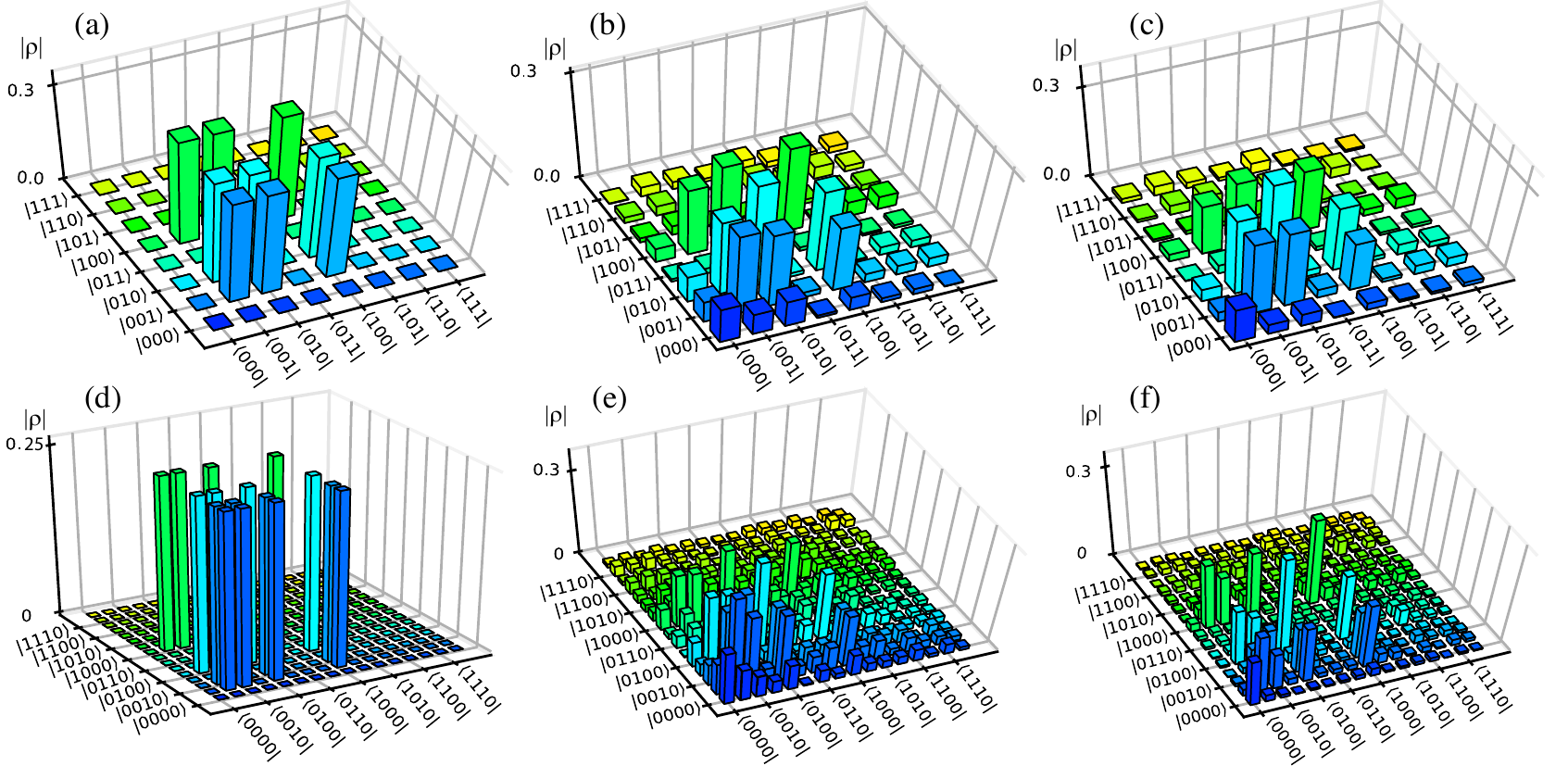}
    \caption{Absolute value of reconstructed density matrix elements of $W_3$ states (upper row) and $W_4$ (lower row). Otherwise same disposal as in Fig.~\ref{fig:GHZ_tomo}.}
    \label{fig:W_tomo}
\end{figure*}

\section{Characterizing $GHZ_N$ and $W_N$ on IBMQ}
\label{sec:charactSec}

Once the entangled states are generated using the improved algorithms described above, it is interesting to characterize the result quality. 

From the quantum measurement principle we cannot determine the state of a system by a single measurement. Even in the classical realm, imagine you want to determine the shape of a real 3D object. In that case, you would need to observe this object from different points of view to reconstruct a full image of it. The complete characterization of a quantum state follows the same principle. For example, measurements in the computational basis (i.e. along the z-axis on the Bloch sphere) cannot differentiate between the states $\ket{0}\pm\ket{1}$, for which both states $\ket{0}$ and $\ket{1}$ are measured half of the time. Complementary measures are needed when no a priori information is known about the states. In the previous example, if we knew that we have one of these two states, a measure along the x-axis of the Bloch sphere would be sufficient to discriminate between these states. 

The tool of choice to characterize a quantum state is called quantum tomography and aims at reconstructing the full density matrix by performing sequences of measurements in different bases on a repeated quantum state. Since the density matrix completely characterizes the state, this method provides maximum information but requires evaluation of $4^N$ expectation values. Each of them requires in turn many repeated measurements.

Quantum tomography is therefore practical only for the lowest values of $N$. In the forthcoming subsection quantum tomography allows to prove that on IBMQ we can achieve $GHZ_N$ and $W_N$ with fidelity $>70\%$ up to $N=5$. For larger $N$ values (up to $N=16$) we must resort to  bypasses presented in the subsequent subsections. For $GHZ_N$ we recourse to decay of parity oscillations (coherence measurement) up to $N=16$, and for $W_N$ to partial tomography and histogram distance. In all cases we systematically find that the performance of logarithmic algorithms is very superior to linear ones.

\subsection{Quantum tomography and fidelity of $GHZ_N$ and $W_N$ for low $N$ values}\label{QtomoAndFSec}

\subsubsection{Simple quantum state tomography}\label{stateTomoSubSec}

State tomography aims at reconstructing the density matrix. For a single qubit, characterized by a $2\times2$ density matrix, the normalized set of Pauli matrices with the identity $\{\frac{1}{\sqrt{2}}I,\frac{1}{\sqrt{2}}\sigma_x,\frac{1}{\sqrt{2}}\sigma_y,\frac{1}{\sqrt{2}}\sigma_z\}$ provides an orthonormal matrix set with respect to the scalar product $\langle A|B\rangle:=\mathrm{Tr}[A^\dagger  B]$. Over this basis the decomposition of the density matrix yields  
$$\rho=\frac{1}{2}\sum_{i=0}^3\langle\sigma_i\rangle\sigma_i$$

\noindent where $\sigma_0$ denotes the identity and $\sigma_{1,2,3}\equiv\sigma_{x,y,z}$. Each coefficient is directly accessed by measuring the operator many times. On IBMQ only one axis, taken as $z$ in the Bloch sphere representation is directly available, so a preliminary rotation is required to measure $\sigma_{x,y}$ (a Hadamard $H$ gate or $S^\dagger H$ gates respectively, where $S\equiv\sigma_z^{1/2}$ is a phase gate). The identity need not to be measured since its expectation value is 1 anyway.

The generalization to $n$ qubits~\cite{Altepeter2004} just involves all the possible tensor product combination of the previous basis set, so the density matrix decomposition reads
$$
\rho=\frac{1}{2^n}\sum_{i_1,...,i_n=0}^3\langle\sigma_{i_1}\otimes\sigma_{i_2}\otimes...\otimes\sigma_{i_n}\rangle \; \sigma_{i_1}\otimes\sigma_{i_2}\otimes...\otimes\sigma_{i_n}
$$ 
The measurements of the $4^n$ real coefficients (sometimes called Stokes parameters by analogy with photonic polarization qubits) itself requires average of many repeated measurements. The default is 1024 shots on IBMQ, but the true limiting factor is the waiting time for processor access. In addition there is a limit to $\sim75$ parallel circuit evaluations (i.e. here different measures), forcing us to split the total work load for $n=5$ into blocks of $64$. To give an idea of the actual resulting limitations, in June 2018 the full quantum tomography of a  $5$-qubit state took about $1~h$, which would increase by a factor $4$ for each additional qubit.

\subsubsection{Reconstructed density matrices}\label{recDMSubSec}

The density matrix reconstructed using the Stokes parameters is most easily represented in the tensor product computational basis. However one should point out that the particular form of our quantum states (in particular our choice of positive phases) ensures that the theoretical value of non-vanishing matrix elements is exactly $+1/2$ for $GHZ_N$ states or $+1/N$ for $W_N$ states. 

The reconstructed matrix elements for $GHZ_3$ and $GHZ_4$ are plotted on Fig.~\ref{fig:GHZ_tomo}. In fact only the absolute value are displayed, so the plots are symmetric around the diagonal. The phases were left aside, because they are noise compared to the characteristic values which are indeed found close to $1/2$. However we can distinguish in the last two columns the effect of $T_1$ relaxation which increasingly favors (with $N$) the overall ground state corresponding respectively to $\ket{0}^{\otimes 3}$ and $\ket{0}^{\otimes 4}$. The non-vanishing off-diagonal elements $|\bra{000}\rho\ket{111}|$ and $|\bra{0000}\rho\ket{1111}|$ are even more strongly affected, which is in fact expected. 

Fig.~\ref{fig:W_tomo} displays similar results for $W_3$ and $W_4$. The plots are a bit more complex, but the component states are easily recognized. The characteristic values are indeed found close to $1/3$ and $1/4$ respectively. Again we can distinguish in the last two columns the increasing effect with $N$ of $T_1$ relaxation towards ground states $\ket{0}^{\otimes 3}$ and $\ket{0}^{\otimes 4}$, and the increasing decay of the off-diagonal elements, as expected. 

\begin{figure}[h]
	\centering
	\includegraphics[width=\linewidth]{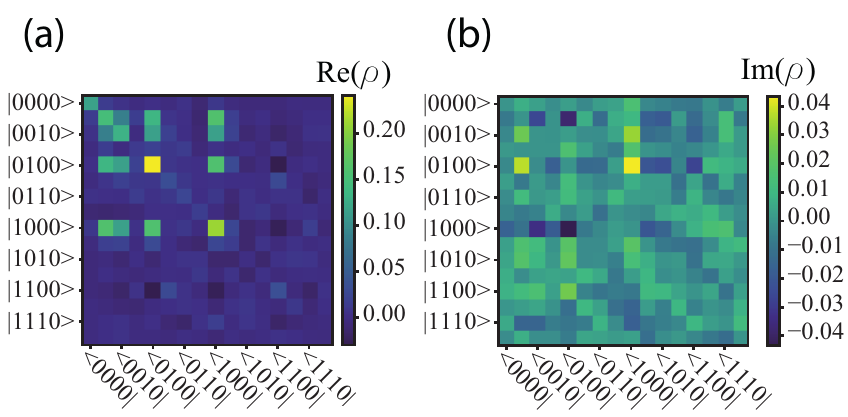}
    \caption{Reconstructed density matrix of $W_4$, with full information about $\textrm{Re}(\rho_{ij})$ and $\textrm{Im}(\rho_{ij})$.}
    \label{fig:fullW4_tomo}
\end{figure}

For completeness we show on Fig.~\ref{fig:fullW4_tomo} the full reconstructed density matrix for $W_4$, with real and imaginary parts instead of absolute values. It proves that the main relevant information was well encoded in the absolute values in Figs.~\ref{fig:GHZ_tomo} and~\ref{fig:W_tomo}.

\begin{figure}[h]
	\centering
	\includegraphics[width=\linewidth]{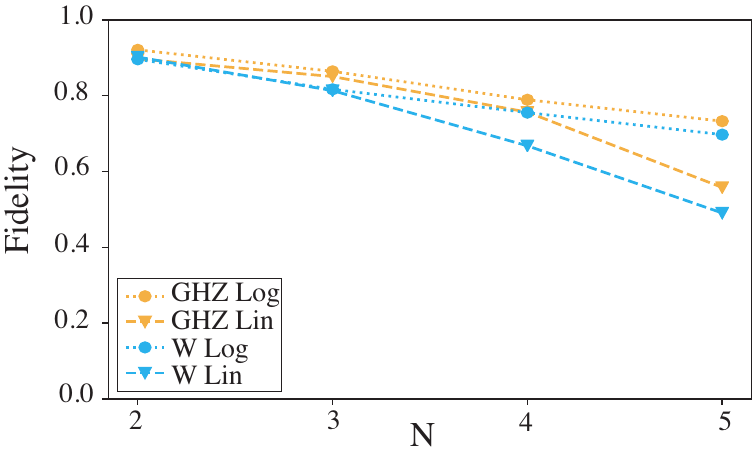}
    \caption{Computed fidelities from the reconstructed density matrices of $GHZ_N$ and $W_N$ states up to $N=5$. Fidelity decays with increasing $N$, but much more slowly for logarithmic algorithms.}
    \label{fig:fidelity}
\end{figure}

These reconstructed density matrices indeed prove that we have generated the desired states up to $GHZ_4$ and $W_4$. Both in Figs.~\ref{fig:GHZ_tomo} and~\ref{fig:W_tomo} the similarity of the plots in the last two columns could lead us to conclude that linear and logarithmic methods produce quantum states of similar quality, which would be surprising. In fact $N=4$ is the turning point where log algorithm starts to have less steps, later on fidelity and histogram distance will reveal this turning point more clearly.

\begin{figure*}[!ht]
	\centering	
    \includegraphics[width=\linewidth]{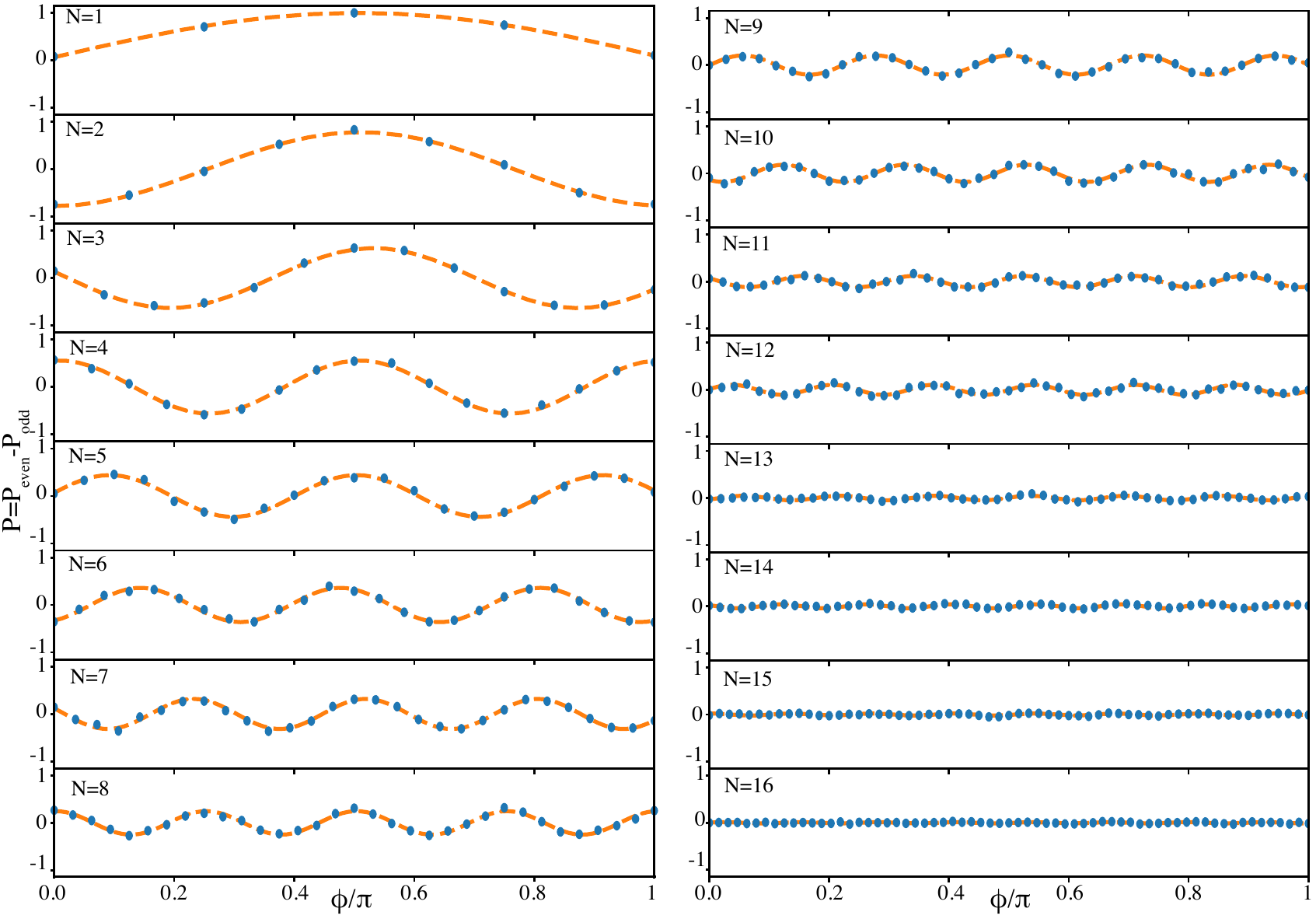}
    \caption{Parity oscillations for each of the $N$-qubit $GHZ$ states for no delay time \(\tau=0\). A single frequency sinusoid, with free amplitude and phase, is fitted to the results for each of the $N$-qubit states (in orange)}
    \label{fig:sinusoids}
\end{figure*}

To be more sensitive than visual inspection we appeal to a distance quantifier between measured density matrix $\rho$ and the ideal $\rho^I$. The fidelity is one such quantifier, and it is defined by~\cite{nielsen_chuang}
$$
F(\rho^I,\rho)=\mathrm{Tr}[\sqrt{\sqrt{\rho^I}\rho\sqrt{\rho^I}}]
$$
which simplifies for pure ideal states into 
$$
F(\psi^I,\rho)=\sqrt{\bra{\psi^I}\rho\ket{\psi^I}}
$$
since in this case $\rho^I=\ket{\psi^I}\bra{\psi^I}$. Fidelity is bounded
$$
0\leq F(\rho^I,\rho)\leq 1
$$
with the nice properties that the lower bound is reached only for perfectly distinguishable states, and that the upper one occurs only when $\rho^I=\rho$. 

In Fig.~\ref{fig:fidelity} we display the computed fidelities from the reconstructed density matrices. We see clearly a decaying fidelity with $N$ for all generated quantum states. Moreover this time the superior characteristics of the logarithmic algorithm are well established, both for $GHZ_N$ and $W_N$ up to $N=5$. We attribute the slightly inferior fidelity reached for $W_N$ versus $GHZ_N$ to the higher number of steps required for $W_N$.

\subsection{Decay of parity oscillations of $GHZ_N$ up to large $N$}

For large $N$ we abandon tomography for parity oscillations' decay in the case of $GHZ_N$.

\subsubsection{Principle of measurement} 

Ideally the density matrix $\rho$ of a $GHZ_N$ state consists of only four elements: two diagonal elements corresponding to the populations of $\ket{0}^{\otimes N}$ and $\ket{1}^{\otimes N}$, as well as two off-diagonal elements corresponding to the relative coherence. The population of the $GHZ_N$ state is then given by $P = \rho_{0\cdots0,0\cdots0} + \rho_{1\cdots1,1\cdots1}$ and its coherence is given by $C = |\rho_{0\cdots0,1\cdots1}| + |\rho_{1\cdots1,0\cdots0}|$. One way to measure the coherence $C$ is by measuring the amplitude of parity oscillations~\cite{Leibfried2005,Monz2011,Ozaeta2017}. The parity is defined as $\mathcal{P} = P_{even} - P_{odd}$ with $P_{even}$ ($P_{odd}$) corresponding to the probability of finding the state with, respectively, an even (or odd) number of excitations. In other words $P_{even}$ ($P_{odd}$) correspond to the probability of finding the state with, respectively, an even (or odd) number of 1's in the measured bitstrings. To induce parity oscillations, one rotates the $GHZ_N$ state with the operator 
\begin{equation*}
\bigotimes^N_{j=1} e^{i\frac{\pi}{4}\sigma_\phi^{(j)}},
\end{equation*}
where $\sigma_\phi^{(j)} = \sigma_x^{(j)} \cos \phi + \sigma_y^{(j)} \sin \phi$, with $\sigma_x^{(j)}$ and $\sigma_y^{(j)}$ the Pauli operators on qubit $j$. This amounts to rotating each qubit of the $GHZ_N$ state by
\begin{equation*}
R(\phi) = \cos \left( \frac{\pi}{4} \right) I + i \sin \left( \frac{\pi}{4} \right) \begin{pmatrix}
   0 & e^{-i\phi} \\
   e^{i\phi} & 0 
\end{pmatrix}
\end{equation*}
When the phase $\phi$ is varied, oscillations of the parity $\mathcal{P}$ are induced. The amplitude of these oscillations directly relates to the coherence $C$.

To experimentally quantify the decoherence of an entangled $GHZ_N$ state, we measure its coherence $C$ as a function of a delay time $\tau$ between the state preparation and the parity oscillations' amplitude measurement. The following method is applied. First, a $\ket{GHZ_N}$ entangled state is generated using our improved algorithms described above. Second, for each $N$, we introduce different delay time $\tau$. Third, for each $\tau$, we vary the angle $\phi$ by applying rotations $R(\phi)$ to each qubits in order to observe parity oscillations. Finally, we measure the $N$ qubits in the computational basis. The amplitude of the measured parity oscillations directly relates to the coherence $C(N,\tau)$ for a $GHZ$ state with $N$ entangled qubits and a delay time $\tau$. The decay of the coherence of the state for different delay times is characterized by the coherence time parameter $T_2^{(N)}$. This method has been previously used by A. Ozaeta and M. McMahon, who measured decoherence in states of up to 8 entangled superconducting qubits~\cite{Ozaeta2017}. Here we present decoherence measurements of entangled systems of up to 16 superconducting qubits.

\subsubsection{Results}

Each circuit is run multiple times in order to decrease statistical errors in the results: 1024 shots until $N=12$ qubits, inclusive, and 8192 onwards. For each $N$-qubit state, $4N+1$ measurements are made with $\phi$ varied from $\phi=0$ to $\phi=\pi$. A sinusoid is fitted to the results, the amplitude of which is related to the coherence $C(N,\tau)$. The results for no delay time ($\tau=0$) are displayed in Figure \ref{fig:sinusoids}. It can be seen that the amplitude of the sinusoid, and thus the coherence of the state, decreases with an increasing number of qubits $N$. However, it is still possible to fit a sinusoid to the results, even for the highest $N=16$ state, as the errorbars are lower due to higher shot count. 

\begin{figure}[h]
	\centering
    \includegraphics[width=\columnwidth]{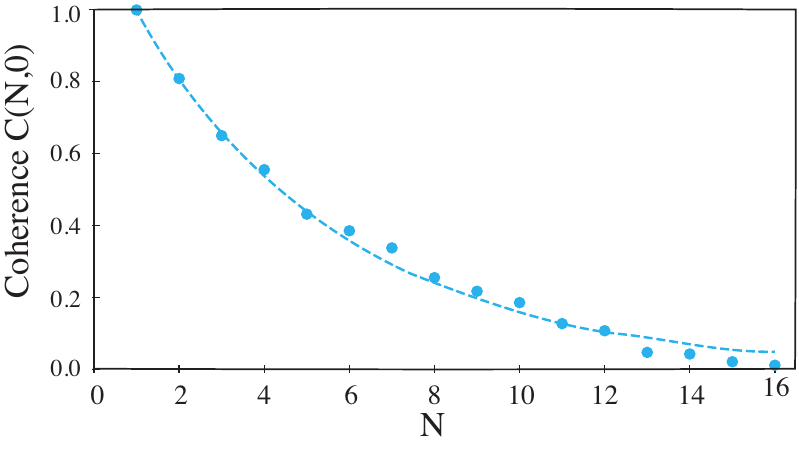}
    \caption{Coherence at no delay $C(N,\tau=0)$ extracted from the parity oscillations as a function of $N$. Exponential decay is observed.}
    \label{fig:CvsN0}
\end{figure}

In Figure \ref{fig:CvsN0}, the coherence of the qubits for no delay $C(N,0)$ is plotted as a function of the number of qubits $N$. The coherence $C(N,0)$ is observed to decay exponentially as a function of $N$, in contrast to the linear decay observed by Ozaeta and McMahon~\cite{Ozaeta2017}. There might be several reasons for this difference. Foremost, their state generation method is $O(N)$, so the time it takes for the state to be created increases linearly. In our case, it is $O(\log_2 N)$ with our efficient algorithm, so the time it takes for the state to be created increases logarithmically. As a first order approximation, we can expect the coherence of the formed state, right after its creation (no delay $\tau=0$), to be proportional to the multiplied error rates of the gates, that is, $C(N,0) \propto E_R^N E_M^{(N-1)}$, where $E_R$ is the average readout error (and there are $N$ readouts) and $E_M$ is the average CNOT gate error (and there are $N-1$ CNOT gates). It should also be proportional to $e^{-t/<T>}$, where $t$ is the time it takes for the qubit to be created and $<T>$ is the average decay rate, associated with the relaxation and decoherence. However, this still slightly overestimates the observed coherence. Additional decay could arise from a progressive loss of coherence of the qubits the longer  it has passed since calibration. Indeed, parameter drift of the qubits occurs throughout the course of a computation~\cite{Tannu2018}. If it is the case then, since the results for low $N$ were obtained first, and closer to the calibration than high $N$, one could expect an extra error affecting more the state the higher $N$ is. Recent work has shown that parameter drift can be compensated for during computation~\cite{Kelly2016}.

\begin{figure}[h]
	\centering
    \includegraphics[width=\columnwidth]{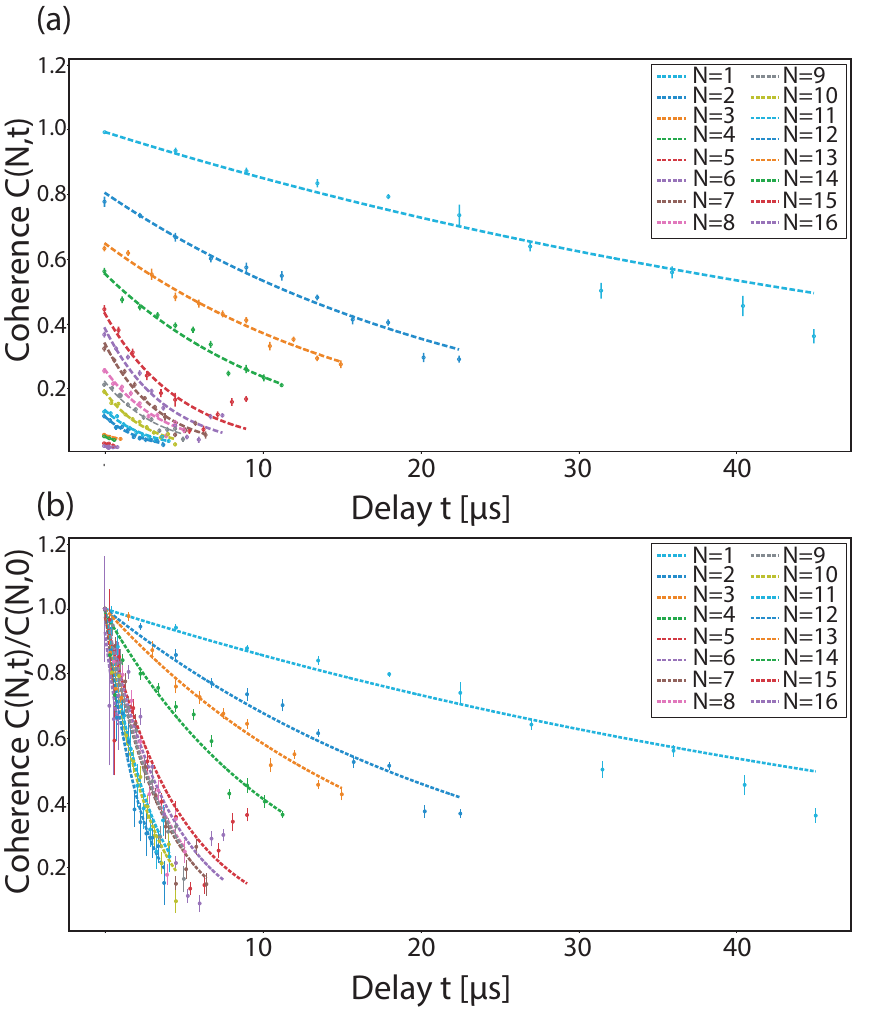}
    \caption{(a) Coherence of each of the generated $GHZ_N$ states plotted against the delay time $\tau$. An exponential function is fitted to the results in order to extract coherence time parameters $T_2^{(N)}$. The error bars correspond to the errors in the coherence value from the fitting of the parity oscillations. Those errors were weighted while fitting the exponential function using the least squares method. (b) Coherence $C(N,\tau)$ normalized to $C(N,0)$ as a function of the delay time $\tau$.}
    \label{fig:Cvstunnorm}
\end{figure}

This coherence measurement is repeated for different delay times, using the sinusoid fit to extract the amplitude, and thus the coherence $C(N,\tau)$ for each $N$-qubit state as a function of the delay time $\tau$. The coherence $C(N,\tau)$ as a function of delay time $\tau$ is displayed in Figure~\ref{fig:Cvstunnorm}(a). An exponential decay is observed. The coherence time parameter $T_2^{(N)}$ characterizing the decay is extracted by fitting an exponential function to the experimental data. Note that our generated states up to 12-qubits have better coherence than the 8-qubit state generated by Ozaeta and McMahon~\cite{Ozaeta2017}. Our efficient algorithm allows a reliable measurement of parity oscillations up to the maximum available number of qubits offered by the ibmqx5 chip, i.e. up to $N=16$. The coherences from 5-qubits onwards seem to be closer in their values. This might be due to a qubit with lower coherence limiting the coherence of the full entangled state, or to the increase in the number of steps to implement the circuit. In fact, the real number of steps increases at $N=5, 8, 10, 13$ and $16$, which, in most cases, is when we see more noticeable drops in coherence.

\begin{figure}[h]
	\centering
    \includegraphics[width=\columnwidth]{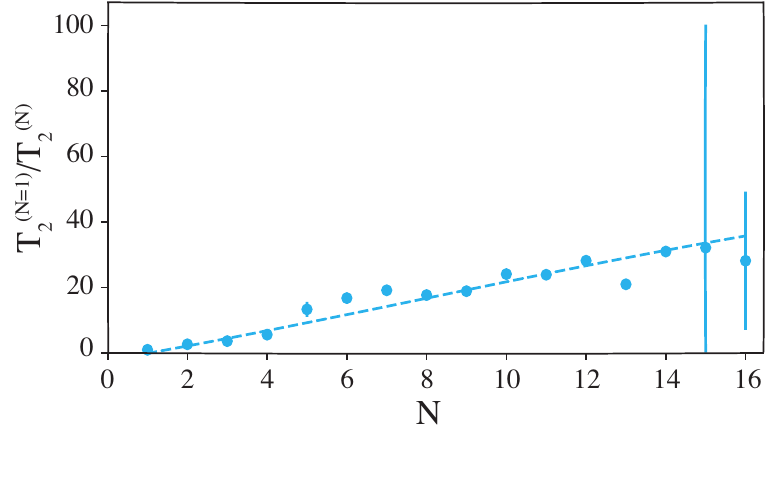}
    \caption{Measured decoherence rate \(1/T_2^{(N)}\) normalized to the decoherence rate for the single qubit state \(1/T_2^{(N=1)}\).}
    \label{fig:T1TN}
\end{figure}

The effect of loss of coherence due to the delay of state generation is traced out by normalizing $C(N,\tau)$ to $C(N,0)$ as displayed in Figure \ref{fig:Cvstunnorm}(b). The decoherence time parameter $T_2^{(N)}$ is also normalized to the single-qubit decoherence parameter $T_2^{(N=1)}$. The results are shown in Figure \ref{fig:T1TN} as a function of $N$. It can be seen that the scaling of the decoherence parameter with the amount of entangled qubits $N$ is linear, and scales approximately with $2N$. The deviation from the linearity can be explained by the fact that not all qubits are equally coherent. For the measurements, the best available qubits are used first, which may influence favorably the decay rates for the lower $N$ states in our experiment. Indeed, if a single-qubit coherence decays proportionally to $e^{-\frac{t}{T_{2,j}}}$  where $T_{2,j}$ is the decoherence time of qubit $j$, the $N$-qubit state coherence would decay proportionally to $\bigotimes_{j=1}^N e^{-\frac{t}{T_{2,j}}}$. Thus one would expect $$\frac{1}{T_2^{(N)}} = \sum_{j=1}^N \frac{1}{T_{2,j}}.$$ At first this can be approximated to $\frac{1}{T_2^{(N)}} \simeq \frac{N}{T_{2}^{(N=1)}}$. In fact, if we assume that the ratio $\frac{T_2^{(N=1)}}{T_2^{(N)}}$ must scale with $N$, and we compute $1/T_2^{(N)}$ vs $N/T_{av}$, with $T_{av}$ a free parameter, we would expect $T_{av}$ to be equal to $T_2^{(N=1)}$. However, we measure that $T_{av} \approx 27 \mu $s, which is a much lower value than 102.2 $\mu$s, the $T_2^{(N=1)}$ value (from the calibration) for the qubit used in the 1-qubit $GHZ$ state. Therefore, the scale factor is understandably higher. If instead of $T_2^{(N=1)}$ we use the average of $N\cdot T_2^{(N)}$, which gives $T^{(N)}_{av} \simeq 25.45 \mu$s, then the ratio scales with $0.943N$. This proves that qubits with low $T_2$ impacts the overall state coherence drastically.

Using our improved algorithms, the decoherence of the $GHZ_N$ states is properly measured as a function of delay time for states with up to $N=16$ qubits, an improvement from 8 entangled qubits by Ozaeta et al.~\cite{Ozaeta2017}. The decoherence rate is found to increase linearly with $N$. Furthermore, the coherence as a function of the number of qubits for zero delay is observed to decay exponentially.

\subsection{Partial tomography of $W(N)$ for large $N$ values: histogram distance from ideal state populations}

$W_N$ states could also be generated for larger $N$ values, up to $N=9$ with linear algorithm and up to $N=16$ with logarithmic algorithm. It is however necessary to evaluate the quality of the result despite the fact that full tomography becomes impractical, as already shown. A natural way out is to restrict the statistical evaluation to state populations (in the computational basis). It should be noted that their ideal values are extremely simple in the case of $W(N)$: only $N$ of them take equal value $1/N$, all the others vanish. We shall call such restricted procedure ``partial tomography'' although there are still $2^N$ populations, which remains a huge quantity of statistical information. 

\begin{figure*}[!t]
	\centering
	\includegraphics[width=\linewidth]{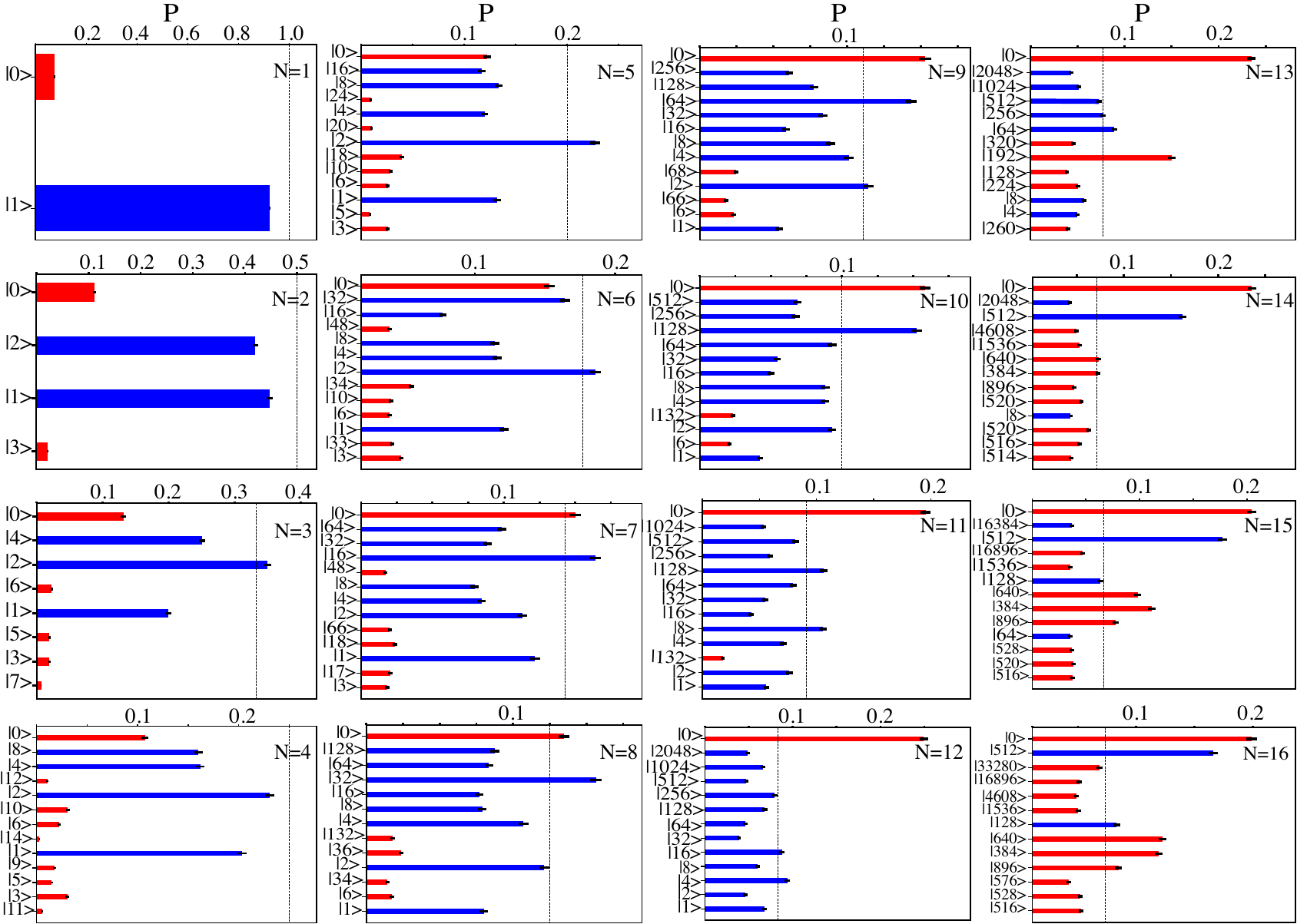}
    \caption{Samples of $W_N$ state populations for logarithmic algorithm (evaluated with 8192 shots), for $N=1$ to $N=16$. In blue all the populations which should reach $1/N$ (dotted line), in red a sample of populations which should vanish. For readability, we denote the states by their equivalent in decimal basis (for instance the state $\ket{010}$ reads $\ket{2}$).}
    \label{fig:W7_partial_tomo}
\end{figure*}

$W(N)$ was generated by the logarithmic algorithm for all values of $N$, and in Fig.~\ref{fig:W7_partial_tomo} we display the 13 most populated basis states together with the measured populations. 8192 shots were used to decrease statistical variance of the populations. States whose target population should in principle be $1/N$ or $0$ are denoted with blue and red lines respectively. $1/N$ target value is indicated by the dotted line.

Visual inspections of Fig.~\ref{fig:W7_partial_tomo} reveal that until $N=12$ blue populations are indeed clearly dominant over the red ones, which indicates generation of fairly good $W_N$ states. However a notable exception is the ground state population corresponding to $\ket{0}^{\otimes N}$ component which rapidly takes significant amplitude as $N$ grows. We can easily explain this fact by relaxation processes which favor final ground state: obviously the longer the state creation, the more important the effect of relaxation. For completeness we also show error bars calculated using the variance for a binomial distribution, so one gets $\sigma=\sqrt{p(1-p)/n}$, where $n$ is the number of shots and $p$ is the average probability associated with the histogram bar.

To have a more objective and finer quality indicator we can use $D_K(N)$ which is defined as the histogram distance (also called Kolmogorov distance, c.f. formula (9.1) of~\cite{nielsen_chuang}) of the $2^N$ state populations with respect to ideal distribution: 
$$
D_K(N)=\frac{1}{2} \sum_{i_1,...,i_n=0}^1 \abs{\bra{i_1,...,i_n} (\rho-\rho^I) \ket{i_1,...,i_n}}
$$
It is easy to compute and has the property that $D_K(N)\rightarrow 0$ indicates highly identical ditributions, whilst $D_K(N)\rightarrow 1$ indicates unrelated distributions.

A more natural measure of distance between the ideal and measured density matrices would be the \textit{trace distance} (c.f.~\cite{nielsen_chuang} Chapter 9, formula (9.11)), but it would require full tomography. Nevertheless one should realize that, when $\rho$ and $\rho^I$ commute, Kolmogorov distance $D_K(N)$ and trace distance are identical~\cite{nielsen_chuang}. Of course we cannot expect that the ideal and measured density matrices would commute perfectly, but still it is a nice property of $D_K(N)$ that for close-to-ideal states the two notions coincide. Nevertheless, by a general theorem (see \cite{nielsen_chuang} theorem 9.1), it should be noted that the trace distance upper bounds the Kolmogorov distance, not the converse, so we cannot directly assert that trace distance is small when Kolmogorov distance is small. 

\begin{figure}
	\centering
	\includegraphics[width=\linewidth]{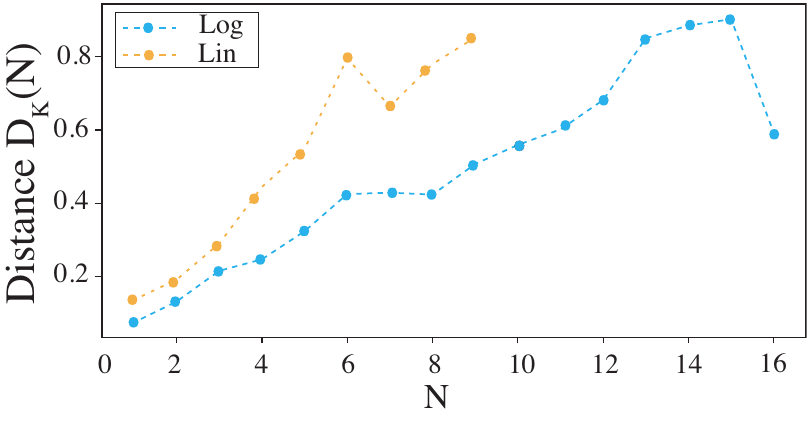}
    \caption{Histogram distance $D_K(N)$ of $W_N$ state populations from ideal, as a function of $N$ for linear and logarithmic algorithms.}
    \label{fig:TV_W}
\end{figure}

Fig.~\ref{fig:TV_W} displays histogram distance $D_K(N)$ for increasing $N$ and for both linear and logarithmic algorithms. First it indicates a poor fidelity for $N>12$. Second, it shows the clear superiority of the logarithmic algorithm over the linear one, for the same number of qubits. This is expected too because detrimental relaxation processes act longer for linear $W_N$ algorithm which involves a bigger number of steps, and the difference increases with $N$. Overall, the histogram distance seems to increase fairly linearly with $N$ for both algorithms, until it saturates at unity, the maximum value.

\section{An attempt at quantum error correction}\label{qerSec}

We implemented the error correction algorithm for Bell and GHZ states presented in \cite[Sections II \& III]{GAPP2017}. The algorithm corrects an arbitrary phase, a phase-flip and a bit-flip error, with the help of {\it ancilla} qubits which store the true phase and parity of the GHZ states. For example, when a GHZ state $\frac{1}{\sqrt{2}} (| 000 \rangle + | 111 \rangle)$ is corrupted to $\frac{1}{\sqrt{2}} ( | 001 \rangle - e^{i \pi / 8}) |110 \rangle )$, arbitrary phase error correction rectifies to $\frac{1}{\sqrt{2}} ( | 001 \rangle - |110 \rangle )$, phase-flip error correction rectifies to $\frac{1}{\sqrt{2}} ( | 001 \rangle + |110 \rangle )$ with one ancilla qubit and bit-flip error correction rectifies to the true state with two ancilla qubits.

\begin{figure}[!hbtp]
\centering
\includegraphics[width=1\linewidth]{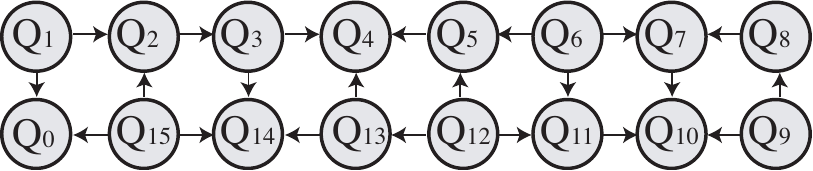}
\caption{Architecture of the 16-qubit Rueschlikon processor}
\label{fig:archi_rueschlikon}
\end{figure}

In our implementation we explore only GHZ states with $N=3$ and $N=4$ so there is not much gain in using the logarithmic circuit instead of linear ones (for $N=3$ they are the same). We have used the linear circuit, at the same time respecting the constraints imposed by the architecture of the 16-qubit machine shown in Figure~\ref{fig:archi_rueschlikon}. A CNOT gate operates on two qubits only if they are connected by an arrow; moreover, an operation opposite to the arrow direction requires a qubit SWAP at the backend of the computer. Such SWAPs must be avoided as much as possible, because their accumulation can lead to non-negligible decoherence. For this purpose we carefully choose the qubits to use:
\begin{itemize}
\item{3-qubit GHZ state: we store the GHZ state in the array LQ=[Q2,Q3,Q4] and the ancilla qubits in LP=[Q15, Q14, Q13]}
\item{4-qubit GHZ state: we store the GHZ state in the array LQ=[Q1,Q2,Q3,Q4] and the ancilla qubits in LP=[Q0, Q15, Q14, Q13]}
\end{itemize}
The arrays LQ and LP start with the index 0.

\subsubsection{Construction of ancilla qubits}

For the ancillas the phase checking circuit (\cite[Fig. 1]{GAPP2017}) requires a series of CNOT gates where a phase ancilla qubit is added on each of the GHZ state qubits. We implement this as follows:
\begin{algorithmic}
\For{$k=N-1$ to $1$}
	\State CNOT(LP[$k$], LQ[$k$])
	\State SWAP(LP[$k$], LP[$k-1$])
\EndFor
\end{algorithmic}
The phase ancilla qubit initially stored in LP[$n-1$] will shift to LP[$0$] after the above loop.

The next parity checking circuit (\cite[Fig. 2]{GAPP2017}) requires another series of CNOT gates where each pair of CNOT gates produce a parity ancilla qubit. This is implemented as follows:
\begin{algorithmic}
\For{$k=0$ to $N-2$}
	\State CNOT(LQ[$k+1$], LP[$k+1$])
	\State SWAP(LP[$k+1$], LP[$k$])
    \State CNOT(LQ[$k$], LP[$k$])
\EndFor
\end{algorithmic}
Eventually, LP[$k$] stores the parity of the qubits (LQ[$k$], LQ[$k+1$]) for $k \in \{0, \dots, N-2\}$, while LP[$N-1$] stores the phase of the GHZ state.

\subsubsection{Circuits for error correction}

We follow the circuits for arbitrary phase error correction, phase flip error correction and bit-flip error correction in Figures 5, 6 and 7 in~\cite{GAPP2017}, respectively. The series of CNOT gates in the circuits of arbitrary phase and phase-flip error correction can be implemented with the same kind of procedures as for phase checking. The series of CNOT gate in the bit-flip error correction circuit is implemented as follows:
\begin{algorithmic}
\For{$k=0$ to $N-2$}
	\State CNOT(LQ[$k$], LP[$k$])
	\State SWAP(LP[$k$], LP[$k+1$])
    \State CNOT(LQ[$k+1$], LP[$k+1$])
    \State CNOT(LP[$k+1$], LQ[$k+1$])
    \State SWAP(LP[$k$], LP[$k+1$])
\EndFor
\end{algorithmic}

\subsubsection{Experimental results}

\begin{figure}[h]
	\centering
    \includegraphics[width=\columnwidth]{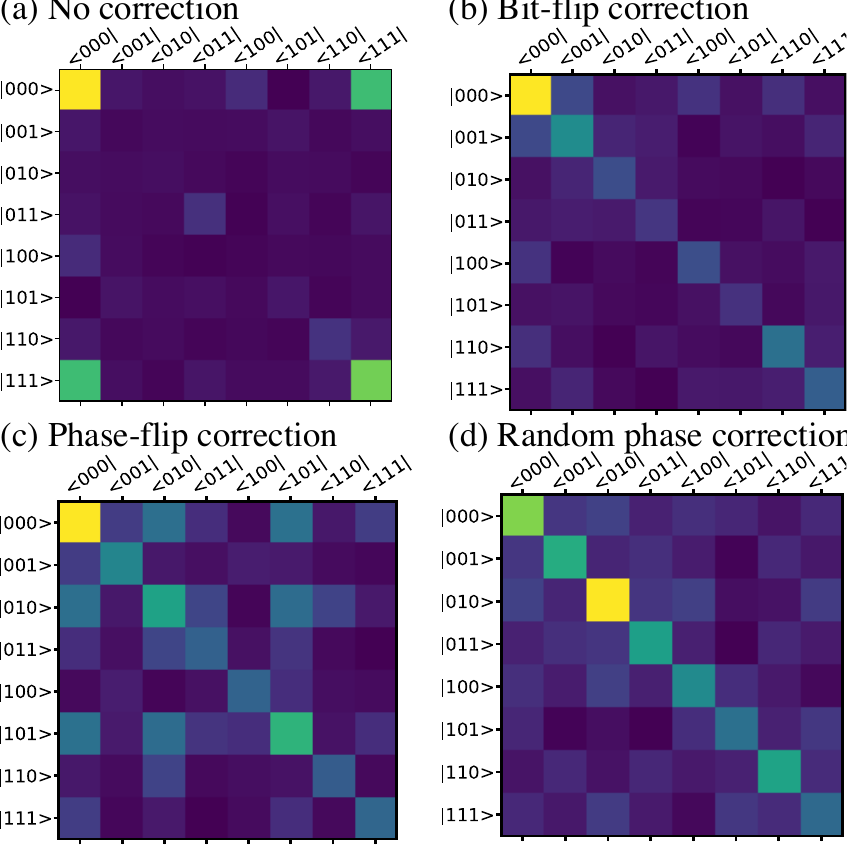}
    \caption{Results from tomography; yellow color indicates higher values and lower values are darker.}
    \label{fig:error_correction}
\end{figure}

We perform tomography on the $GHZ_3$ state (based on the method outlined in the previous section) without and with error correction. We manually initialize the ancilla qubits so that only the error correction circuits are verified. 1024 shots are executed for each value of the coefficients in the density matrix expansion. No time-lapse or delays are added between state creation, error correction and tomography. The results for bit-flip, phase-flip and arbitrary phase correction are shown in Figure~\ref{fig:error_correction}, where the absolute values of the matrix elements of density matrices are plotted.

One observes a general decay of the GHZ states after the error corrections in all three figures. In addition, there is a general relaxation towards $|000\rangle\langle000|$ (apart from the random phase correction in Fig.\ref{fig:error_correction}(d)). 

In order to further test the effect of decoherence, one can compare the reconstructed density matrix without any error correction but after a time-lapse (or a delay) of roughly equivalent to the number of gates used in the error correction circuit; preliminary results seem to indicate that the amount of decoherence is much smaller. We do not believe this is a problem of the error correction scheme per se. Rather, the noise level in the experiment is too high and creates many errors well above the error correction capability of the proposed scheme.\\


\section{Conclusion}

In conclusion, we have proposed and implemented efficient algorithms with logarithmic step complexities for the generation of entangled $GHZ_N$ and $W_N$ states on IBMQ. These algorithms are of high relevance for quantum networks too since they are potential working horses to create deterministically a shared $GHZ_N$ or $W_N$, this without any prior entanglement, and even if the number of parties is {\em arbitrary}. This has many promising applications~\cite{Yesilyurt2016}. We have generated and characterized these entangled states with up to $N=16$ superconducting qubits on the IBM quantum computer, twice the number of qubits previously achieved. Different characterization methods were discussed: full quantum tomography for low-$N$ GHZ and W states, parity oscillations and histogram distance for large $N$ GHZ and W states respectively. They prove the efficiency and robustness of our algorithms, and reveal the current limitations of the IBM Q platform in terms of decoherence. Quantum error correction on the generated GHZ states was attempted and proved to be limited by the current level of decoherence. The observed  linear increase in decoherence rate with number of qubits is promising given the latest improvement in experimental quantum computing~\cite{Neill2018}. New processors such as the QS1\_1 or ibmq\_20\_tokyo will be available commercially from IBM in 2018, and will offer superior coherence properties and higher connectivity. They should allow to further demonstrate the superior performance of the logarithmic algorithms presented here. Overall we have experienced that the \textit{Open Source Quantum Information Science Kit} (QISKit)~\cite{QISKit} enables the deployment of large experiments on currently available quantum computers.

\section*{Acknowledgment}

We acknowledge use of the IBM Quantum Experience for this work. The views expressed are those of the authors and do not reflect the official policy or position of IBM or the IBM Quantum Experience team.

\newpage

\subsection{Appendix: logarithmic construction of $\ket{W_{11}}$}

For general $N$ the dichotomy tree for the logarithmic construction of $\ket{W_N}$ can be formalized as follows:

\begin{algorithmic}[1]
\State We start with the couple $(\lfloor N/2 \rfloor, N)$ at the root
\For{every generation}
\For{every leaf $(n,m)$ not equal to (0,1), (1,1) or (1,2)}
	\State Perform the dichotomies $n = \lfloor n/2 \rfloor + \lceil n/2 \rceil$ and $m=\lfloor m/2 \rfloor + \lceil m/2 \rceil$
    \State Leaf $(n,m)$ generates the upper child $(\lfloor n/2 \rfloor, \lfloor m/2 \rfloor)$ and the lower child $(\lceil n/2 \rceil, \lceil m/2 \rceil)$
\EndFor
\EndFor
\For{every pair of children which contains $(1,1)$}
	\State Swap the pair of children
\EndFor
\State Eliminate all $(0,1)$ and $(1,1)$
\end{algorithmic}

\begin{figure}[!hbtp]
\begin{subfigure}{0.25\textwidth}
\includegraphics[width=\linewidth]{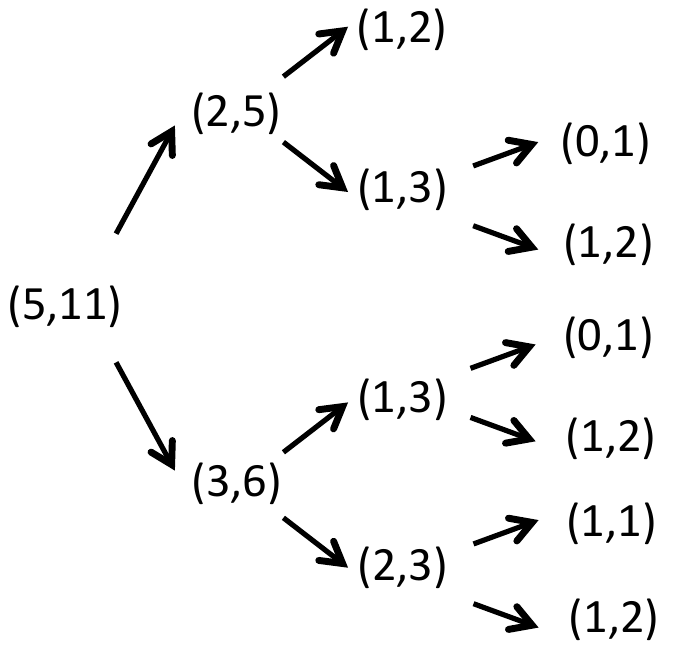}
\caption{Basic construction}
\end{subfigure}\hspace*{\fill}
\begin{subfigure}{0.25\textwidth}
\includegraphics[width=\linewidth]{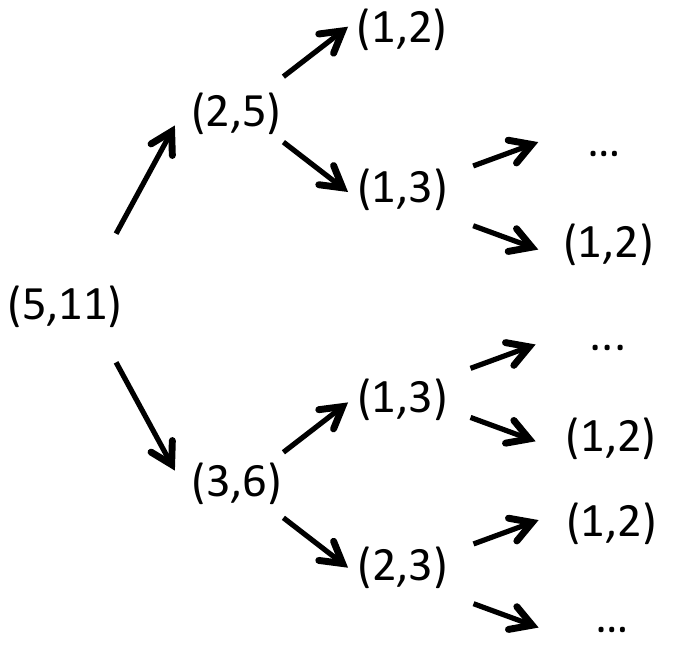}
\caption{After pruning}
\end{subfigure}
\caption{Dichotomy tree for the construction of $\ket{W_{11}}$}
\label{figs:logW-tree-N=11}
\end{figure}

\begin{figure}[!hbtp]
	\centering
	\includegraphics[width=\linewidth]{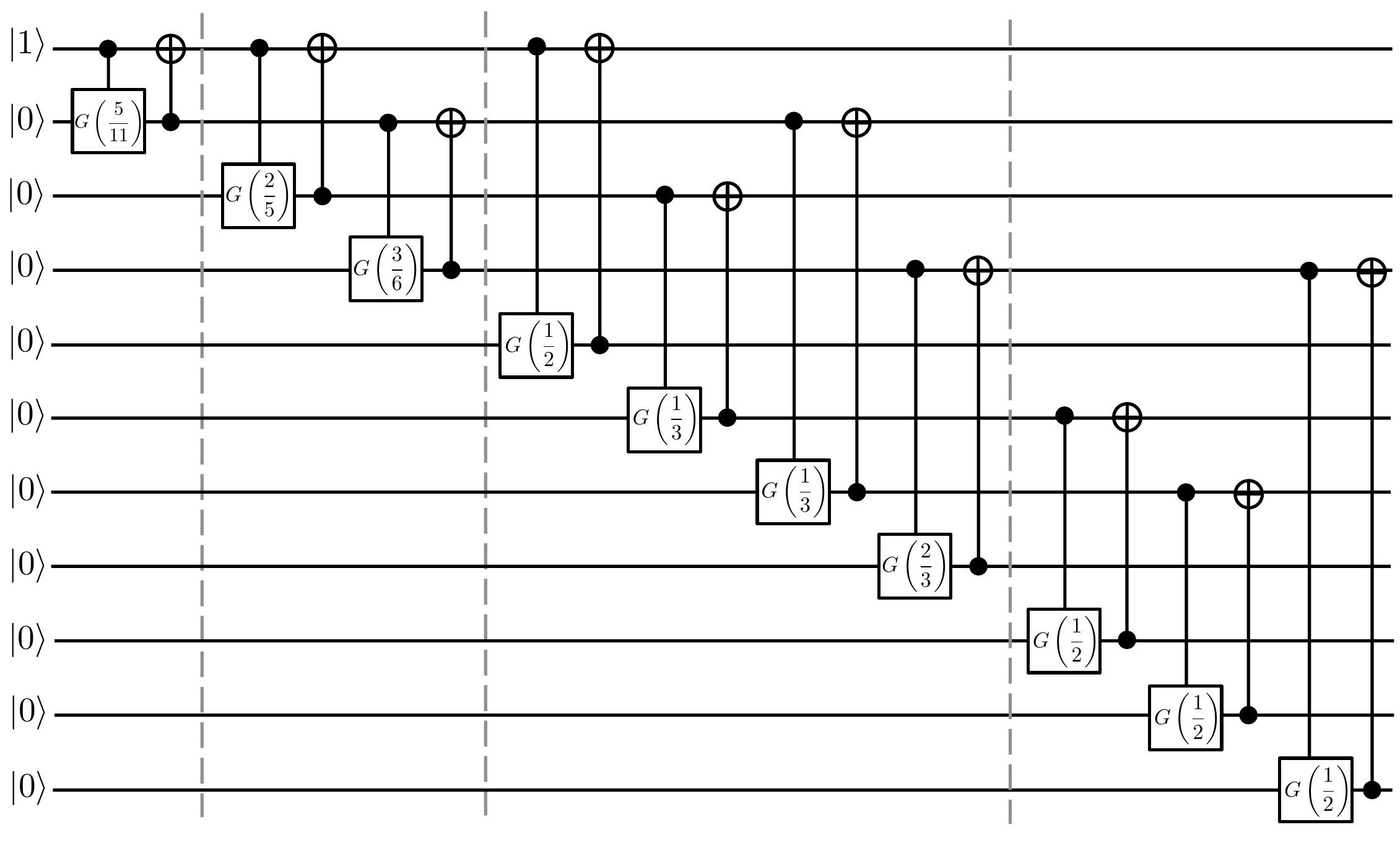}
    \caption{Logarithmic complexity circuit for $\ket{W_{11}}$}
    \label{figs:logW-N=11}
\end{figure}

Let us apply the algorithm to $N=11$.
The root is $(\lfloor N/2 \rfloor, N) = (5,11)$. Lines 2 to 7 construct a tree illustrated in Fig.~\ref{figs:logW-tree-N=11}a. Then lines 8 to 11 decorate the tree and output the final version in Fig.~\ref{figs:logW-tree-N=11}b. The circuit in Fig.~\ref{figs:logW-N=11} is then built by connecting the successive block according to the tree structure on the r.h.s.

Again we provide the intermediate states at the end of each time slice for the reader's convenience. At the end of the first time slice the state is
\begin{align*}
\big(\sqrt{\frac{5}{11}} \ket{10} + \sqrt{\frac{6}{11}} \ket{01} \big) \otimes \ket{000000000}
\end{align*}
At the end of the second time slice the state is
\begin{align*}
& \big ( \sqrt{ \frac{5}{11}\frac{2}{5}} \ket{1000} + \sqrt{\frac{5}{11}\frac{3}{5}} \ket{0010} + \sqrt{ \frac{6}{11}\frac{3}{6}} \ket{0100} \\
& \hspace{0.5cm} + \sqrt{\frac{6}{11}\frac{3}{6}} \ket{0001} \big ) \otimes \ket{0000000}
\end{align*}
At the end of the third time slice the state is
\begin{align*}
& \big ( \sqrt{\frac{5}{11}\frac{2}{5}\frac{1}{2}} \ket{10000000} 
+ \sqrt{\frac{5}{11}\frac{2}{5}\frac{1}{2}} \ket{00001000} \\
&  + \sqrt{ \frac{5}{11}\frac{3}{5}\frac{1}{3}} \ket{00100000} + \sqrt{ \frac{3}{5} \frac{5}{11}\frac{3}{5}\frac{2}{3} } \ket{00000100} \\
&  + \sqrt{\frac{6}{11}\frac{3}{6}\frac{1}{3}} \ket{01000000} + \sqrt{ \frac{3}{6} \frac{6}{11}\frac{3}{6}\frac{2}{3} } \ket{00000010} \\
&  + \sqrt{\frac{6}{11}\frac{3}{6}\frac{2}{3}} \ket{00010000} + \sqrt{\frac{6}{11}\frac{3}{6}\frac{1}{3}} \ket{00000001} \big ) \otimes \ket{000}
\end{align*}
Finally the circuit output is
\begin{align*}
& \sqrt{\frac{5}{11}\frac{2}{5}\frac{1}{2}} \ket{10000000000} 
+ 
\sqrt{\frac{5}{11}\frac{2}{5}\frac{1}{2}  } \ket{00001000000} \\
&  + \sqrt{ \frac{5}{11}\frac{3}{5} \frac{1}{3}} \ket{00100000000} 
+ \sqrt{ \frac{5}{11}\frac{3}{5} \frac{2}{3} \frac{1}{2}} \ket{00000100000} \\
& 
+ \sqrt{ \frac{5}{11}\frac{3}{5} \frac{2}{3} \frac{1}{2}} \ket{0000000100} 
+ \sqrt{\frac{6}{11}\frac{3}{6}\frac{1}{3}} \ket{01000000000} \\
& 
+ \sqrt{\frac{6}{11}\frac{3}{6}\frac{2}{3}\frac{1}{2}} \ket{00000010000} 
+ \sqrt{\frac{6}{11}\frac{3}{6}\frac{2}{3}\frac{1}{2}} \ket{00000000010} \\
& 
+ \sqrt{\frac{6}{11}\frac{3}{6}\frac{2}{3}\frac{1}{2}} \ket{00010000000} 
+ \sqrt{\frac{6}{11}\frac{3}{6}\frac{2}{3}\frac{1}{2}} \ket{00000000001} \\
& 
+ \sqrt{\frac{6}{11}\frac{3}{6}\frac{1}{3}} \ket{00000001000} = \ket{W_{11}}
\end{align*}

\end{document}